%% file: draft.tex
\documentclass[journal,onecolumn]{IEEEtran}

\usepackage{booktabs}
\usepackage{subcaption}
\usepackage{float}
\usepackage{cite}
\usepackage{amsmath}
\usepackage{algorithmic}
\usepackage{array}
\usepackage{url}
\usepackage{float}

\DeclareMathOperator*{\argmin}{arg\,min}

\newcount\Comments  
\usepackage{color}
\definecolor{darkgreen}{rgb}{0,0.5,0}
\definecolor{purple}{rgb}{1,0,1}
\definecolor{orange}{rgb}{1,0.5,0}
\definecolor{cyan}{rgb}{0.7,0.7,0}
\definecolor{rust}{rgb}{0.1,0.7,0.7}
\definecolor{gold}{rgb}{0.7,0.5,0.1}

\ifCLASSINFOpdf
   \usepackage[pdftex]{graphicx}
\else
\fi

\begin{document}

\title{I-24 MOTION: \\An instrument for freeway traffic science}




%
%
%

\author{
    \IEEEauthorblockN{Derek Gloudemans\IEEEauthorrefmark{1}\IEEEauthorrefmark{3}, Yanbing Wang\IEEEauthorrefmark{2}, Junyi Ji\IEEEauthorrefmark{2}, Gergely Zachar\IEEEauthorrefmark{2}, Will Barbour\IEEEauthorrefmark{2}, Daniel B. Work\IEEEauthorrefmark{1}\IEEEauthorrefmark{2}\\}
    \IEEEauthorblockA{\IEEEauthorrefmark{1}Vanderbilt University Department of Computer Science\\}
    \IEEEauthorblockA{\IEEEauthorrefmark{2}Vanderbilt University Department of Civil and Environmental Engineering\\}
    \IEEEauthorblockA{\IEEEauthorrefmark{3}derek.gloudemans@vanderbilt.edu}
} 

\maketitle

\begin{figure}[!b]
    \centering
    \includegraphics[width = \textwidth]{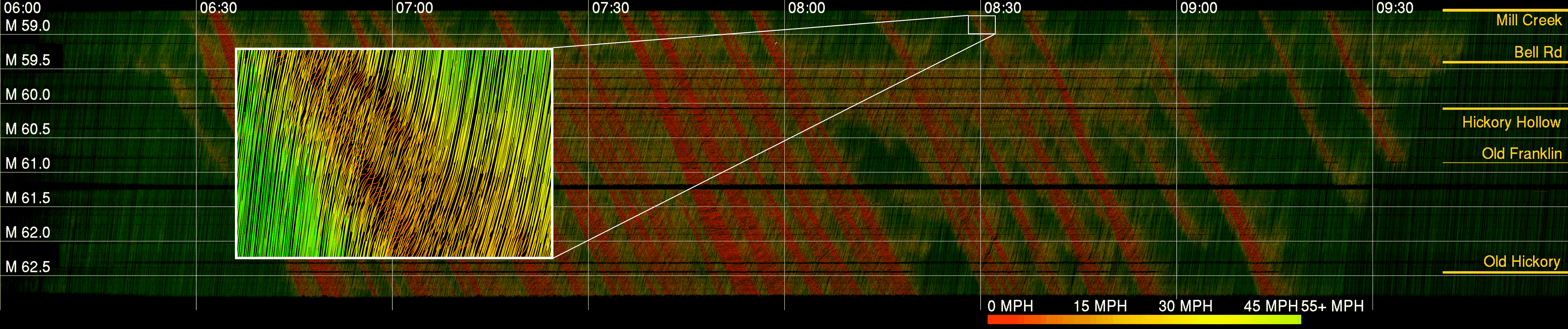}
    \caption{Time-space diagram for four hours of I-24 W morning rush hour traffic on Nov 25, 2022, generated from I-24 MOTION vehicle trajectories. x-axis: time of day (HH:MM); y-axis roadway postmile (mi). Postmile decreases for travelers in the westbound direction. A typical congestion pattern is shown with frequent oscillatory traffic observed; and recurring waves travel upstream relative to the direction of traffic at 12-13 mph. The names of interchanges and overpasses appear on the right. The figure inset shows a zoomed in portion of the data which is 0.25 mi in length and 4 min in duration.}
    \label{fig:teaserTimeSpaceDiagram}
\end{figure}

\begin{abstract}
The Interstate-24 MObility Technology Interstate Observation Network (I-24 MOTION) is a new instrument for traffic science located near Nashville, Tennessee.  I-24 MOTION consists of 276 pole-mounted high-resolution traffic cameras that provide seamless coverage of approximately 4.2 miles I-24, a 4-5 lane (each direction) freeway with frequently observed congestion. The cameras are connected via fiber optic network to a compute facility where vehicle trajectories are extracted from the video imagery using computer vision techniques. Approximately 230 million vehicle miles of travel occur within I-24 MOTION annually. The main output of the instrument are vehicle trajectory datasets that contain the position of each vehicle on the freeway, as well as other supplementary information, vehicle dimensions, and class. This article describes the design and creation of the instrument, and provides the first publicly available datasets generated from the instrument. The datasets published with this article contains at least 4 hours of vehicle trajectory data for each of 10 days. As the system continues to mature, all trajectory data will be made publicly available at \url{i24motion.org}.
\end{abstract}


%


\section{Introduction}

Transportation science is undergoing a digital transformation in which increasingly automated vehicles are being developed and deployed on roadways, changing the fundamental physics of traffic flow. Even a small number of automated vehicles can have a direct impact on the macroscopic behavior of traffic flow, highlighting the need to monitor and observe traffic flows across microscopic and macroscopic scales. 

At the same time new vehicles are being introduced that may alter the flow, new technologies are advancing that ease the ability to capture the behavior of traffic at scales that were impossible to realize even a few years ago. For example, automated vehicles now transmit critical contextual data about the surrounding environment on the vehicle Controller Area Network (CAN), allowing opportunities to measure vehicle spacings and relative velocities, which were not possible using only GPS devices in phones and vehicles. Drone technologies have reached a degree of maturity that now facilitate camera based monitoring over roadways at impressive spatial scales. While these advancements offer opportunities to accelerate traffic flow science, there are still direct needs for monitoring the individual and collective behavior of vehicles over long temporal and spatial resolutions.

Recognizing the impact of freeway trajectory data collection efforts such as NGSIM~\cite{alexiadis2004next} and HighD~\cite{krajewski2018highd} (see also Table~\ref{tab:data-comparison}), and emerging urban datasets exemplified by pNEUMA~\cite{barmpounakis2020new}, and at the same time the limited availability of sources for trajectory data, we started on a 5-year effort to instrument a section of freeway that could help enable the next wave of empirical traffic science that depends on abundant trajectory datasets. This article presents the outcome of that effort, resulting in an instrument known as I-24 MOTION. 

I-24 MOTION is a camera-based trajectory generation system located on I-24 near Nashville, TN. The instrument consists of 276 4K resolution video cameras mounted on 40 poles ranging from 110 ft to 135 ft above the freeway. The cameras are positioned with overlapping fields of view and are connected by a fiber optic network to a compute facility where the videos are converted to vehicle trajectories. The instrument captures approximately 230 million vehicle-miles of travel annually, and experiences regular recurring congestion. 

Figure~\ref{fig:teaserTimeSpaceDiagram} illustrates the data captured by I-24 MOTION, showing a time-space diagram spanning 4.2 miles of I-24 westbound traffic during 4 hours of morning congestion starting at 6:00AM. The image is created by plotting all westbound vehicle trajectories and color-coding the points based on the speed of the vehicle. Vehicle lengths, widths, heights, and lateral positions are also measured but not shown. The waves visible in the image propagate at approximately 12-13 miles per hour. Data used to generate this diagram are released with this work.

The main contribution of this article is the creation of the I-24 MOTION instrument, which generates the trajectory datasets released with this work. The article provides the description of key elements of the instrument, including the road network geometry and features, the features of the cyber-physical assets that compose the instrument, and the general data processing steps. It also shares initial datasets and introduces the location for where future datasets will be released. 

These elements of this article are critical to understand the uses and limitations of the current and future datasets. For example, as we explain in Section~\ref{sec:system}, the cameras are pole-mounted. The height of the poles are selected to minimize occlusion (excellent for generating accurate vehicle trajectories), but the height can allow sway in strong winds (bad for generating accurate vehicle trajectories). Thus, the physical design directly influences the types of artifacts that can be introduced. The datasets released by I-24 MOTION will be provisioned with a digital object identifier and change logs as new data processing algorithms are deployed and as artifacts are removed. 

We also provide a preliminary description of the datasets, the known artifacts today, and our plans to improve them over time. It is clear that at a macroscopic scale, the data in the initial release can already support novel macroscopic analysis and insight, since no interpolation is required - all 4 miles are observed. At the same time, we describe known issues (e.g., fragmented trajectories due to tracking failures; fragmented trajectories due to a vehicle crash which damaged hardware on one pole, etc.). Some of these issues will be resolved through instrument maintenance cycles; while others will be resolved with the advancement of better automated data generation methods. As individual datasets mature, and new datasets are introduced, this article will serve as the reference point for users of all future datasets generated by the instrument.

\begin{table*}[t]
\centering
\begin{tabular}{llllrrrr} 
\toprule 
\textbf{Dataset} & \textbf{Location} & \textbf{Context} & \textbf{Year} & \textbf{Cameras} & \textbf{Time Scale} & \textbf{Spatial Scale}  & \textbf{Vehicles} \\
\midrule
NGSIM US-101\cite{alexiadis2004next} & Los Angeles, CA & 5-6 lane highway & 2005 & 8 & 0.75 hr   & 0.64 km &  9,206  \\
HighD \cite{krajewski2018highd} & Cologne, GE & 2-3 lane highway & 2018 & 1 & 16.5 hr & 0.42 km & 110,500 \\
ExiD \cite{exiDdataset} & Aachen and Cologne, GE & 2-4 lane interchanges & 2021 & 1 & 16.1 hr & 0.42 km & 69,172 \\
Automatum \cite{spannaus2021automatum}  & GE & 2-4 lane highway & 2021  & 1 & 30 hr & 0.66 km & 60,000 \\
HIGH-SIM \cite{shi2021video} & I-75, FL & 3-4 lane highway & 2021 & 3 & 2 hr & 2.44 km & -\\
Zen Traffic Dataset \cite{seo2020evaluation} & Osaka, JP & 2 lane highways & 2018 & - & 5 hr & $\sim$2 km & - \\
\midrule 
I24-MOTION (released) & Nashville, TN & 4-5 lane highway & 2022 & 276 & 47 hr & 6.75 km & $\sim$600,000 \\
I24-MOTION (planned) & Nashville, TN & 4-5 lane highway & 2023 & 276 & daylight & 6.75 km & $\sim$150,000/day \\
\bottomrule
\end{tabular}
\caption{Comparison of existing highway complete vehicle trajectory datasets. ``$\sim$'' indicates approximate value. ``-'' indicates data is not available. }
\label{tab:data-comparison}
\end{table*}

The remainder of this article is organized as follows: Section \ref{sec:review} reviews the literature landscape around vehicle trajectory data, situating this work among existing research efforts.  Section \ref{sec:system} describes the physical infrastructure, hardware, and software systems of I-24 MOTION. Section \ref{sec:data} describes the data produced in more detail, including the recorded quantities, coordinate system, and a comparison in spatio-temporal scale to existing vehicle trajectory dataset. Section \ref{sec:future} provides some preliminary analysis of the data including a characterization of the wave propagation speeds observed in the datasets. Lastly, Section \ref{sec:conclusion} highlights the future direction of the instrument. 

\section{Related Work} 
\label{sec:review}

\subsection{Data collection for traffic modeling}

At the macroscopic level, traffic phenomena are often observed and described with three quantities of interest, i.e., flow, speed, and density \cite{may1990traffic}. Fundamental diagrams~\cite{turner201175} like the Greenshields and Greenberg models~\cite{greenshields1935study,greenberg1959analysis} relate the traffic quantities while models such as the Lighthill-Whitham-Richards (LWR) \cite{lighthill1955kinematic} and the Aw–Rascle–Zhang (ARZ) ~\cite{aw2000resurrection} are developed to describe the spatio-temporal evolution of traffic. These models can be validated with data collected from  radar-based devices and loop detectors \cite{roess2004traffic}. Large-scale macroscopic data monitoring systems such as the freeway performance measurement system (PeMS)~\cite{choe2002freeway}
 in the United States; the A5 freeway near Frankfurt \cite{schonhof2007empirical} in Germany; and the M42 highway \cite{stewart2006highways} in England; and later floating-vehicle measurement-based on cell phone carrier data \cite{bar2007evaluation} or GPS positional data \cite{herrera2010evaluation} have enabled research on macroscopic traffic flow dynamics \cite{helbing1997empirical,helbing1998jams,kerner1999physics,treiber2000congested,zheng2011applications}. A challenge is that the data typically must be interpolated spatially (in the case of inductive loops), or scaled up across all vehicles (in the case of probe data) to gain a complete spatio-temporal picture.

 Unlike the accumulated average macroscopic data and models, microscopic models give attention to the  interactions between individual vehicles. Since the early car-following experiments~\cite{chandler1958traffic} conducted by physically connecting vehicles to measure space gap, many emerging in-vehicle technologies including on-board radar detectors \cite{kesting2008calibrating}, cameras \cite{jones2001keeping}, laser sensors \cite{gohring2011radar} and global positioning system (GPS) devices~\cite{gurusinghe2002multiple,ma2006estimation} have been applied to measure vehicle spacing, speed and relative speed. 
 
 With the advances in visual sensing, video-based trajectory data from road-side cameras, high buildings, helicopters and drones gradually has become a mainstream source for microscopic modeling \cite{treiterer1974hysteresis,alexiadis2004next,ossen2005car,ossen2006interdriver,ossen2008validity,krajewski2018highd}. Trajectory data with the complete information for specific road segments supported a range of efforts including the development, calibration and validation of  car-following models \cite{tordeux2010adaptive,koutsopoulos2012latent}, lane-change modeling, trajectory prediction \cite{deo2018multi,altche2017lstm}, and traffic oscillation  analysis~\cite{yeo2009understanding}. 

Some traffic phenomena benefit from observation of traffic across the micro and macroscopic scales. For example, traffic waves observable at the macroscopic scale can result from instabilities and disturbances in the flow at the level of individual vehicles~\cite{treiterer1974hysteresis,laval2006lane}. Macroscopic data, frequently used for traffic wave studies, can cover a great spatiotemporal scale that reveals the dynamics of traffic waves on road networks, but it is unable to provide insight into why the wave is generated and how it is propagated. Trajectory data can help provide these insights \cite{li2014stop,laval2010mechanism,zheng2011applications} when available with adequate spatiotemporal coverage. 
Hence, abundant trajectory datasets, as highlighted in the article~\cite{li2020trajectory}, can enable traffic research at both the macroscopic and microscopic scales, aiding in understanding traffic phenomena like jam clusters and state transition dynamics \cite{schonhof2007empirical,kerner2005physics,seo2017traffic}. It can also capture the complex interaction within multiple-class traffic participants for heterogeneous traffic flow \cite{khan1999modeling,arasan2005methodology,ambarwati2014empirical}.

\subsection{Existing Testbeds}
I-24 MOTION also operates as an open road testbed, which allows experiments to be conducted on the freeway and measured using the instrument. Existing \textit{closed course} and~\textit{open road} testbeds already address some critical emerging research needs~\cite{emami2020review}. Closed course testbeds, such as the American Center for Mobility \cite{acm2021}, MCity \cite{briefs2015mcity}, GoMentum Station \cite{cosgun2017towards}, and Suntrax \cite{heery2017florida}, have the distinct advantage of being capable of hosting experiments and data collection for cutting edge technologies and techniques including those under active research and development. By testing in highly controlled settings, they can assure safety and eliminate external factors such as unpredictable drivers and road conditions that can confound experiments. Because of the motivating objectives of closed course testbeds, they can be limited in their ability to test in real traffic conditions with regular drivers encountered on public roads. 
Open road testbeds exist in many forms on a variety of road types; examples include the Minnesota Traffic Observatory \cite{parikh2014implementation}, The Ray \cite{ray2021}, the California Connected Vehicle Test Bed \cite{farrell2015precision}, Ann Arbor Connected Vehicle Test Environment \cite{aacvte2021}, and Providentia \cite{krammer2019providentia}. They support experiments in live traffic, similar to the I-24 instrument. Currently, the collection of high-fidelity trajectory data on each and every vehicle on the roadway over a multi-mile scale does not exist the United States, though the Lower Saxony testbed and the Zen Traffic initiative support these objectives in Germany and Japan. Table \ref{tab:testbeds} summarizes these existing vehicle testbeds. 

\begin{table*}[!ht]
\centering
\begin{tabular}{lllllr} 
\toprule 
\textbf{Testbed} & \textbf{Location} & \textbf{Sensors} & \textbf{Type} & \textbf{Intended Usage}  \\
\midrule 
ACTION \cite{ACTION2022}                     & Tuscaloosa, AL      & DSRC, Cameras & Open road & CV, V2I \\
M-City \cite{briefs2015mcity}                     & Ann Arbor, MI       & DSRC, Cameras & Closed course & AV   \\
The Ray \cite{ray2021}                    & Interstate 85, GA   & DSRC & Open road & CV, V2I\\
California CV Testbed \cite{farrell2015precision}         & Palo Alto, CA       & DSRC & Open road & CV, V2I\\ 
Gomentum \cite{cosgun2017towards}                   & Concord, CA         & LIDAR, DSRC, Cameras & Closed course & CV, AV\\
ACM Proving Grounds \cite{acm2021}        & Ypsilanti, MI       & DSRC & Closed course & AV \\
SunTrax \cite{heery2017florida}                    & Orlando, FL         & DSRC & Open road & V2I\\
AACTVE \cite{aacvte2021}                    & Ann Arbor, MI       & DSRC & Open road & V2I\\
\midrule
Providentia \cite{krammer2019providentia} & Munich, DE  & Radar,Cameras & Open road & Trajectories \\
Minnesota Traffic Observatory \cite{parikh2014implementation} & Minneapolis, MN & Radar & Open road & Trajectories \\
Lower Saxony Testbed \cite{von2021creating}       & Braunschweig, DE    & LIDAR, DSRC, Cameras & Open road & Trajectories, CV, AV\\ 
Zen Traffic Roadways \cite{seo2020evaluation}           & Osaka, JP           & Cameras              & Open road & Trajectories, CV, AV \\
\midrule
I-24 MOTION   & Nashville, TN   & Cameras & Open road & Trajectories, CV, AV \\
\bottomrule \\
\end{tabular}
\caption{Existing vehicle testbeds. \textit{DSRC} indicates direct short range communications, \textit{Trajectories} indicates complete vehicle trajectory generation, \textit{CV} indicates connected vehicle testing, \textit{V2I} indicates vehicle to infrastructure testing, and \textit{AV} indicates autonomous vehicle testing.}
\label{tab:testbeds}
\end{table*} 
\newpage

\subsection{Emerging Observation Technologies}
In a parallel thread, significant research has been devoted to the computer vision tasks of \textit{object detection} (locating relevant objects within an image) and \textit{object tracking} (associating distinct objects in video frames across time). Especially in the past 10 years, rapid progress has been made in the use of modern hardware \cite{krizhevsky2012imagenet}, neural network architectures \cite{he2016deep,redmon2016you,girshick2015fast, duan2019centernet}, and massive-scale image datasets \cite{deng2009imagenet,lin2014microsoft} to fit  accurate object detection algorithms. Approaches for extracting vehicle trajectory data utilizing these techniques have been proposed. For example, the work \cite{dubska2014automatic} proposes a method to detect vehicle 3D rectangular prism bounding boxes using background subtraction and blob segmentation, relying on automatic parameter extraction of the scene homography proposed in \cite{dubska2014fully}. The work \cite{sochor2018boxcars} uses this data to train a \textit{convolutional neural network} (CNN) to produce the same data without the need for scene-wide calibration. In \cite{ren2018learning}, 2D object detectors are used to estimate vehicle positions on the road plane (the ambiguity of vehicle position within a 2D bounding box is not fully addressed). \cite{subedi2019development} uses ground plane projection of vehicle pixels from multiple cameras to estimate the vehicle's position, validating with turning movement counts. Other solutions rely on re-identification of 2D tracked objects, without addressing 2D annotation position ambiguity \cite{tang2018single,chen2019multi}. Other methods utilize instance segmentation networks \cite{zhao2019real, he2017mask} on traffic scenes with little occlusion. A few approaches \cite{zhang2019longitudinal,malinovskiy2009video} avoid object detection by measuring object presence in \textit{longitudinal scanlines} along each roadway lane, but occlusion and lane changes pose difficult challenges in this problem formulation. In theory, such methods promise to address the shortage of trajectory data.

These advances, along with the increasing prevalence of aerial drones, have enabled recent research efforts to revisit the task of vehicle trajectory extraction and make marked advancements to the state of the art. The HighD, \cite{krajewski2018highd}, ExiD \cite{exiDdataset}, AUTOMATUM \cite{spannaus2021automatum}, and HIGH-SIM \cite{shi2021video} datasets all utilize aerial imagery shot from either drone or helicopter-mounted cameras to produce complete highway vehicle trajectory data, and the \textit{Third Generation Simulation} (TGSIM)~\cite{tgsim} is a similar in-progress effort designed to capture trajectory data containing deployed automated vehicle technologies. Similarly, the pNEUMA \cite{barmpounakis2020new}, inD \cite{inDdataset}, rounD \cite{rounDdataset}, OpenDD \cite{breuer2020opendd}, Interaction \cite{zhan2019interaction} and CitySim \cite{zheng2022citysim} datasets utilize drones or swarms of drones to study complex urban vehicle and pedestrian interactions in more detail. High aerial fields of view make modern image segmentation algorithms \cite{he2017mask} well posed for vehicle tracking in these contexts, but these methods are temporally limited by the relatively short battery life of drones (generally under an hour) and the requirement for human pilots.

\section{System Description}
\label{sec:system}
\begin{figure*}
    \centering
\includegraphics[width=\linewidth]{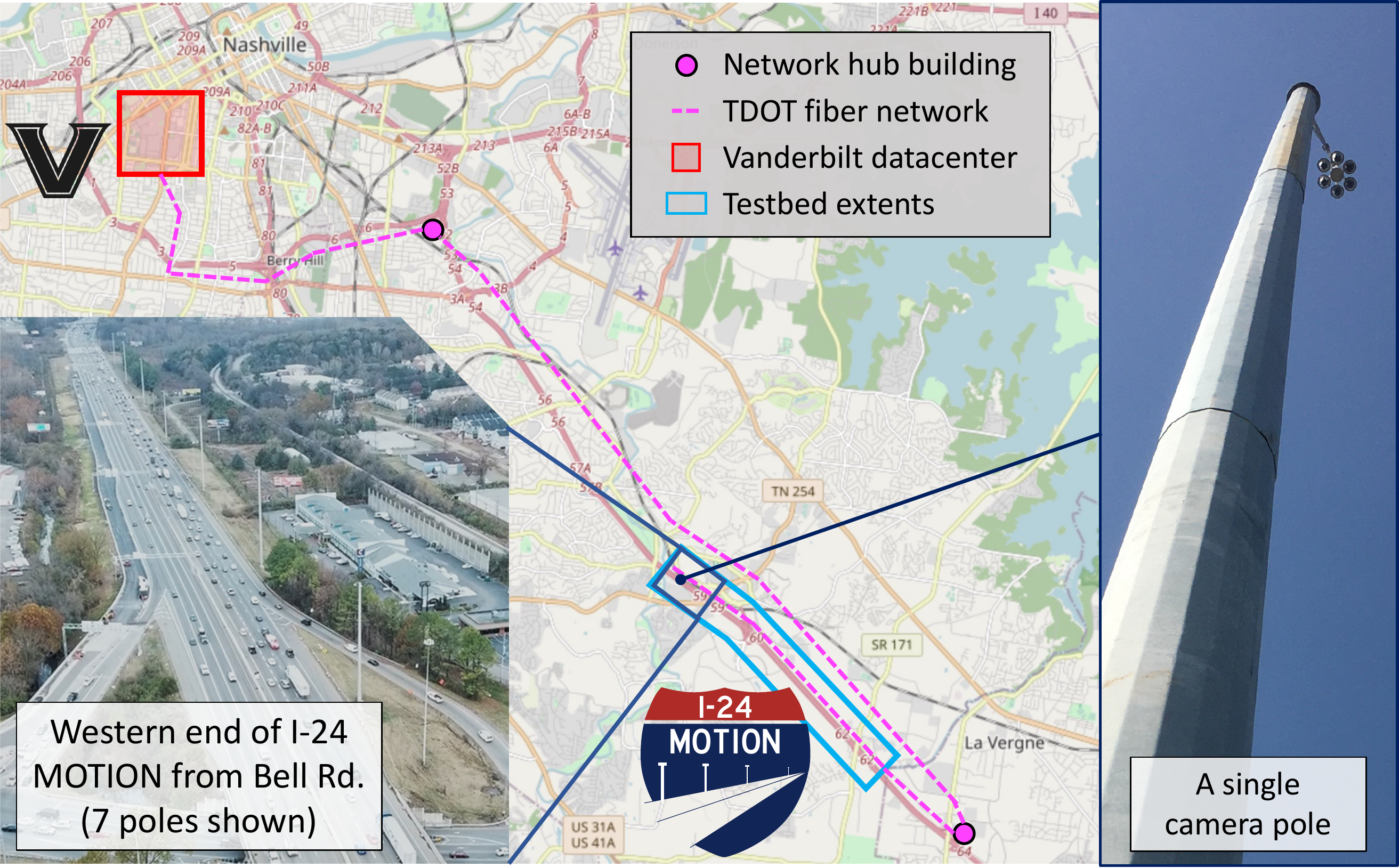}
    \caption{Overview of I-24 MOTION site showing location relative to Nashville, TN. The major TDOT fiber network elements and their connection to Vanderbilt University, which houses the trajectory generation algorithms that operate on the live video feeds, are also shown on the map.}
    \label{fig:location}
\end{figure*}


This section describes the I-24 MOTION instrument, detailing the physical infrastructure, network and compute hardware, and core algorithms required to provide accurate and complete vehicle trajectory data across a large spatial and temporal scale. The system is still in active development, and continual improvements to improve the reliability, accuracy, and processing speed of the system will be made over the following years.


\begin{figure*}[!ht]
    \centering
\includegraphics[width=\linewidth]{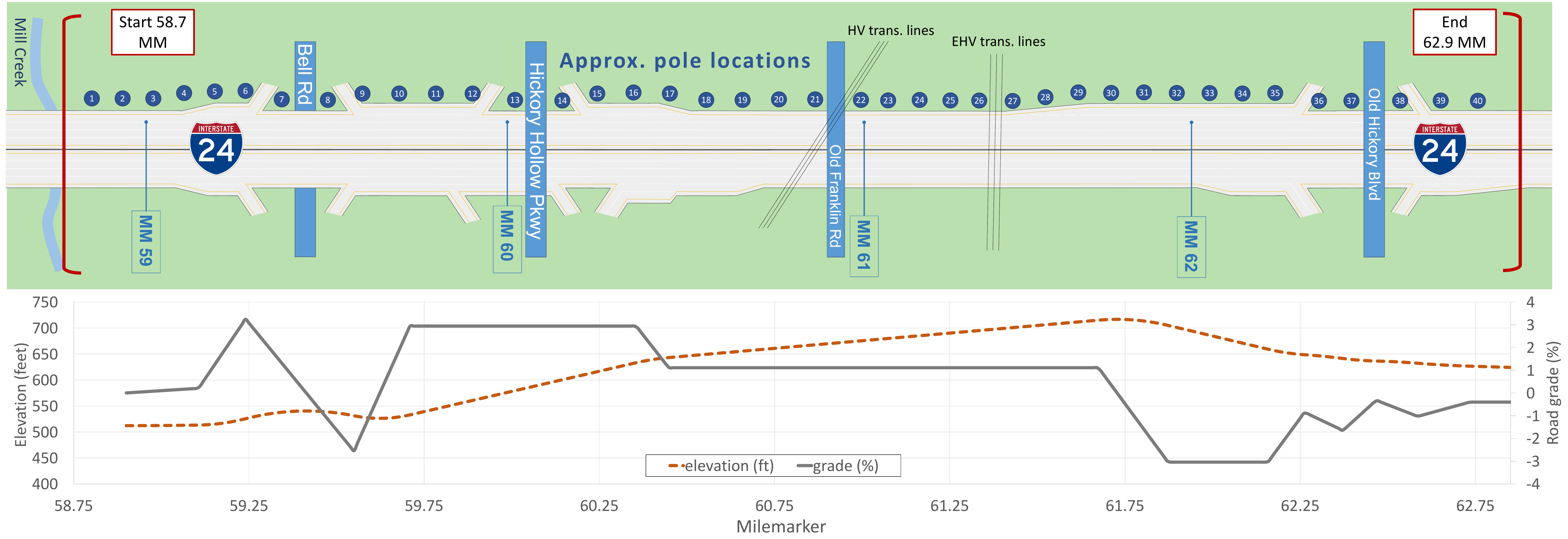}
    \caption{\textbf{TOP:} Diagram of the I-24 MOTION instrument, spanning from Mill Creek (postmile 58.8) to milemarker 62.8. Camera poles (blue circles) are spaced at roughly 550-foot intervals and are shown in the image relative to other key infrastructure elements. The system spans three overpasses and one underpass, as well as three interchanges with 13 entrance/exit ramps. Relative positions of all elements are correct but diagram is not drawn to scale. \textbf{Bottom:} Elevation and road grade along the freeway. Grade is measured in the eastbound (diagram left to right) direction.}
    \label{fig:roadway}
\end{figure*}

\subsection{Physical Infrastructure}
The I-24 MOTION instrument provides a continuous field of view of 4.2 miles (6.75 km) on the 4-5 lanes (each direction) I-24 freeway, southeast of Nashville, Tennessee, USA. Pole mounted cameras are connected via a fiber network to a data center, where computer vision tracking and trajectory processing takes place. A total of 276 4K resolution cameras are mounted on 40 poles, each 110-135 feet tall, spaced every 500-600 feet along the freeway. The poles provide an overhead vantage point of the road to reduce occlusion, and to provide overlapping fields of view. (A separate 3-pole, 18 camera validation system \cite{barbour2021interstate} is located about 0.75 miles eastbound on I-24 from the primary instrument and was used for technology testing and system planning).

\subsubsection{Location}
The location for I-24 MOTION was selected based on traffic conditions, constructability factors, and co-location with other \textit{Tennessee Department of Transportation} (TDOT) initiatives. The four mile section of Interstate 24 is located ten miles southeast of Downtown Nashville and exhibits an \textit{annual average daily traffic} (AADT) of approximately 150,000 vehicles per day across its length~\cite{tdot2022counts}. Morning and afternoon rush hour traffic exhibits reliably heavy congestion in opposite directions, frequently reaching stop-and-go conditions, with easily-observable traffic waves on a typical day. I-24 near Nashville is a heavy commuter and freight corridor (10-15\% of the vehicle traffic are heavy trucks): it links smaller cities of Murfreesboro, La Vergne, and Smryna with Nashville, and serves as a major shipping and industrial transportation route for Middle Tennessee and the southeast United States.


This section of the I-24 corridor was also selected for the state’s first \textit{Integrated Corridor Management} (ICM) project, called the I-24 SMART Corridor, which operates on the 28-mile route between Nashville and Murfreesboro. The ICM project includes Interstate 24, the parallel arterial route SR 1, and connector routes between I-24 and SR 1. The ICM project has deployed an upgraded communications network and Intelligent Transportation System (ITS) devices, such as variable speed limit control, lane control, and ramp metering, for increased operational management of the corridor. This collocation will eventually allow the study of a variety of implemented ITS solutions associated with the I-24 SMART Corridor using I-24 MOTION \cite{chen2015variable, papageorgiou2003review}, when the active traffic management systems are enabled.

\subsubsection{Camera poles}
The 40 I-24 MOTION camera poles are each composed of a steel pole structure, ground-level communications and power cabinet, camera lowering device at the top of the pole, and custom camera cluster assembly, each detailed below. Figure \ref{fig:components} shows select system components. Camera pole locations, as well as various other landmarks of interest, are included in Appendix \ref{app:locations}.

The camera pole system was prototyped across three years at existing pole locations on the TDOT network and with a purpose-built three-pole validation system constructed in 2020 \cite{gloudemans2020interstate, barbour2021interstate}. Valuable lessons from the validation system regarding camera selection, camera cluster mounting position, and pole-to-pole spacing were incorporated in the full system design. The details of the pole components are as follows:

\begin{figure*}[!hb]
    \centering
    \includegraphics[width=\linewidth]{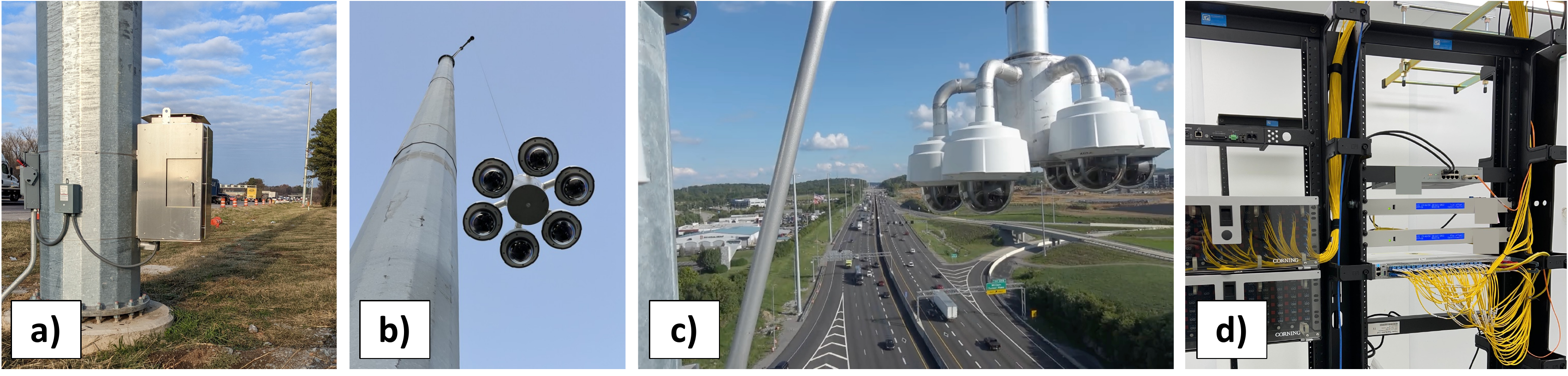}
    \caption{Testbed components: a) view of a camera pole base showing the electrical disconnect, transformer, and ground cabinet; b) camera cluster in the process of lowering to the ground with the CLD; c) view of camera cluster at the top of a pole; d) fiber optic junction at network hub building and GPS network time servers.}
    \label{fig:components}
\end{figure*}

\begin{itemize}
    \item \textbf{Steel pole structure:} To observe all vehicles on the roadway with minimal occlusion, the poles are significantly taller than standard 30-50 ft poles used on many other CCTV systems. New poles and corresponding foundations were designed and built to a standard that the total deflection at the top of the pole is less than 1.5 inches in a 30 mph wind. Average pole-to-pole spacing is 550 feet across the instrument, with a minimum of 425ft and a maximum of 625ft due to roadside obstacles and entrance/exit ramps.
    \item \textbf{Ground-level cabinet:} A pole-mounted cabinet (shown in Figure~\ref{fig:components}a) houses a network switch for the fiber optic network, a fiber patch panel, and two power supplies. Power supplies in the cabinet provide DC power to the fiber network switch in the cabinet (65W supply) and to the camera cluster at the top of the pole (240W supply). A fiber communications backbone is present throughout the instrument and links each pole to a communications hub building (shown in Figure~\ref{fig:components}d) and the rest of the TDOT network. Each pole maintains a one gigabit per second network link to an aggregation network switch in a star network topology. This network topology helps simplify configuration and troubleshooting and has additional resilience in the case of some physical damage scenarios.
    \item \textbf{Camera lowering device:} The \textit{camera lowering device} (CLD) is a critical component of all traffic monitoring cameras in the instrument. It allows the camera cluster to be safely lowered to the ground (see Figure~\ref{fig:components}b) for routine cleaning and maintenance using a winch at the base of the pole. While typically configured for only a single camera on a CLD, manufacturer collaboration and internal bench testing confirmed that the lowering device could support simultaneous data transmission from six 4K resolution video cameras to the ground-level cabinet where it ties into the fiber network. The CLD also contains redundant ethernet and power connections that can be utilized without the need for physical access to the top of the pole in case of a connector failure. The CLD is mounted to the top of the pole with a 54-inch extension arm and angled support strut (shown in Figure~\ref{fig:components}c) for added rigidity.
    \item \textbf{Camera cluster assembly:} Mounted on each pole is a custom, 6-camera mount attached to the camera lowering device (shown in Figure~\ref{fig:components}c). The orientation of the camera cluster is orthogonal to the roadway direction(s) of travel. The weather-tight camera mount holds a network switch which aggregates six video data streams to transmit them through a single gigabit ethernet connection on the CLD. The network switch receives DC power from the pole cabinet and supplies power over ethernet (PoE+) to each of the six cameras at 25.5W. On the six poles adjacent to the three interchanges within the instrument, a second camera cluster assembly is mounted in an orientation pointing towards the under/overpass; in the future these cameras will support trajectory generation for vehicles as they enter and exit the highway.
\end{itemize}

\subsubsection{Video cameras}

\begin{figure*}[b]
    \centering
\includegraphics[width=\linewidth]{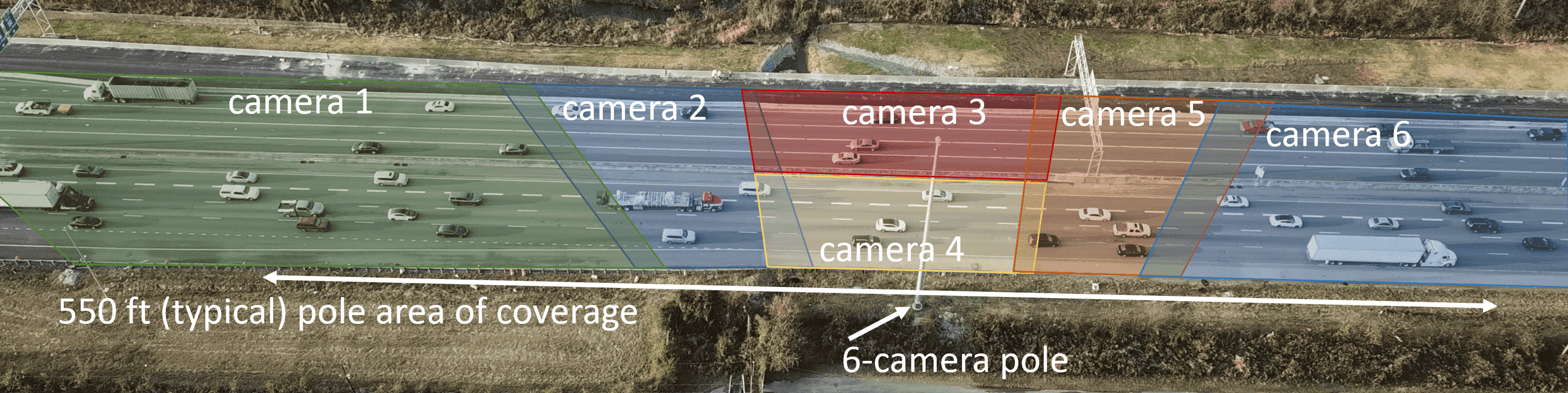}
    \caption{Example camera fields of view for a single 6-camera pole. Each portion of the roadway is covered by at least one camera, with overlaps long enough to allow objects to be tracked between cameras.}
    \label{fig:fov}
\end{figure*}

The cameras on the instrument are a 4K resolution pan/tilt/zoom (PTZ) network IP model, powered by power over ethernet. The PTZ capabilities allow remote alignment to achieve the necessary 180-degree overlapping field of view across cameras on each pole, as seen in Figure~\ref{fig:fov}, and between camera poles. Deploying multiple cameras to each pole extends coverage of the instrument and reduces the number of poles needed. While cameras with wider image field of view exist, these suffered in testing from distortion at the edges of the image that could not easily be corrected to the accuracy needed for coordinate localization.

A critical technical consideration with network IP cameras is time synchronization across cameras and true frame capture time reporting. Cameras are synchronized over \textit{network time protocol} (NTP) to a primary and secondary stratum 1 GPS-based time servers on the local network (in the network hub building) and frequently re-synchronize (roughly every 15 minutes). The camera firmware provides timestamps associated with video frames corresponding at about 10 microsecond  accuracy relative to the camera clock time.  Cameras capture up to 4K resolution video at 30 frames per second. Frame-to-frame timing is typically observed to be uniform (33.3 ms), but in some cases  non-negligible time differences result from duplicated or skipped frames (an artifact of camera exposure requirements as implemented in camera firmware.) Although the camera clocks are are precisely synchronized and the exact frame capture times are different for all devices, accurate time-stamping of each frame allows  processing algorithms to compensate for the relative time offsets for each camera.

\subsection{Network and Compute Hardware Architecture} \label{sec:compute-arch}

All video data feeds are received from the TDOT network into a Vanderbilt data center for processing across a dedicated 40 gigabit fiber network connection. Centralized computing in a data center provides the computing hardware with dedicated, long-term support and infrastructure, in addition to future expansion possibilities. Two network switches support the cluster of servers: a data layer switch with two 25 gigabit connections to each server and a management layer switch for 1 gigabit user connectivity, control, and IPMI. A system control server directs the processing functions of the cluster across ten or more servers/nodes. It hosts a control interface where system managers dispatch processing jobs and propagates job configurations to each node. Nine processing nodes are dedicated to computer vision tracking and the initial trajectory construction. Each node contains eight graphics processing units (GPUs) that decode incoming video and perform object detection and tracking tasks. The nodes track vehicles across cameras, but each node operates independently with statically-assigned cameras. A vehicle traversing the entire instrument will generate a partial trajectory fragment on each of the (nine) processing nodes. Incoming video is buffered on its respective compute node and discarded after processing. Following initial trajectory generation, a post-processing server performs the complete trajectory assembly and reconciliation tasks. The cluster contains two data storage arrays responsible for storing the resulting trajectories -- both initial trajectory fragments and post-processed complete trajectories -- as well as log messages, monitoring data, algorithm training data, and instrument experiment data. Additionally, two servers within the cluster serve as a development and testing environment for new software versions and one server performs ancillary tasks such as large-scale visualization and traffic analysis. 

\subsection{Software Architecture} \label{sec:software}

A prototype software architecture comprises of three main modules: video ingest, vehicle detection and tracking, and trajectory post-processing and reconstruction, managed by the system control server. Before a run session starts, related configuration files and metadata are registered and stored in database for record-keeping or re-processing.

\subsubsection{Video ingestion and recording} The cameras produce a H.264 encoded video, currently at 1080p resolution and 30 frames per second to reduce the data size. The streams are split into 10 minute chunks and recorded into a Matroska (MKV) container. The timestamps, corresponding to the exact exposure moment of each frame (streamed separately in a custom field) which are incorporated into the PTS (Presentation timestamp) metadata during recording. This field is mandatory for video files, thus providing a standardized method for frame timing information, and enables interoperability with any conforming software. The video stream, with the current configuration and all 276 cameras, occupies $\sim$1 TB for each recorded hour at 1080p resolution.

\subsubsection{Vehicle Detection and Tracking}
\begin{figure}
    \centering
    \includegraphics[width=\textwidth]{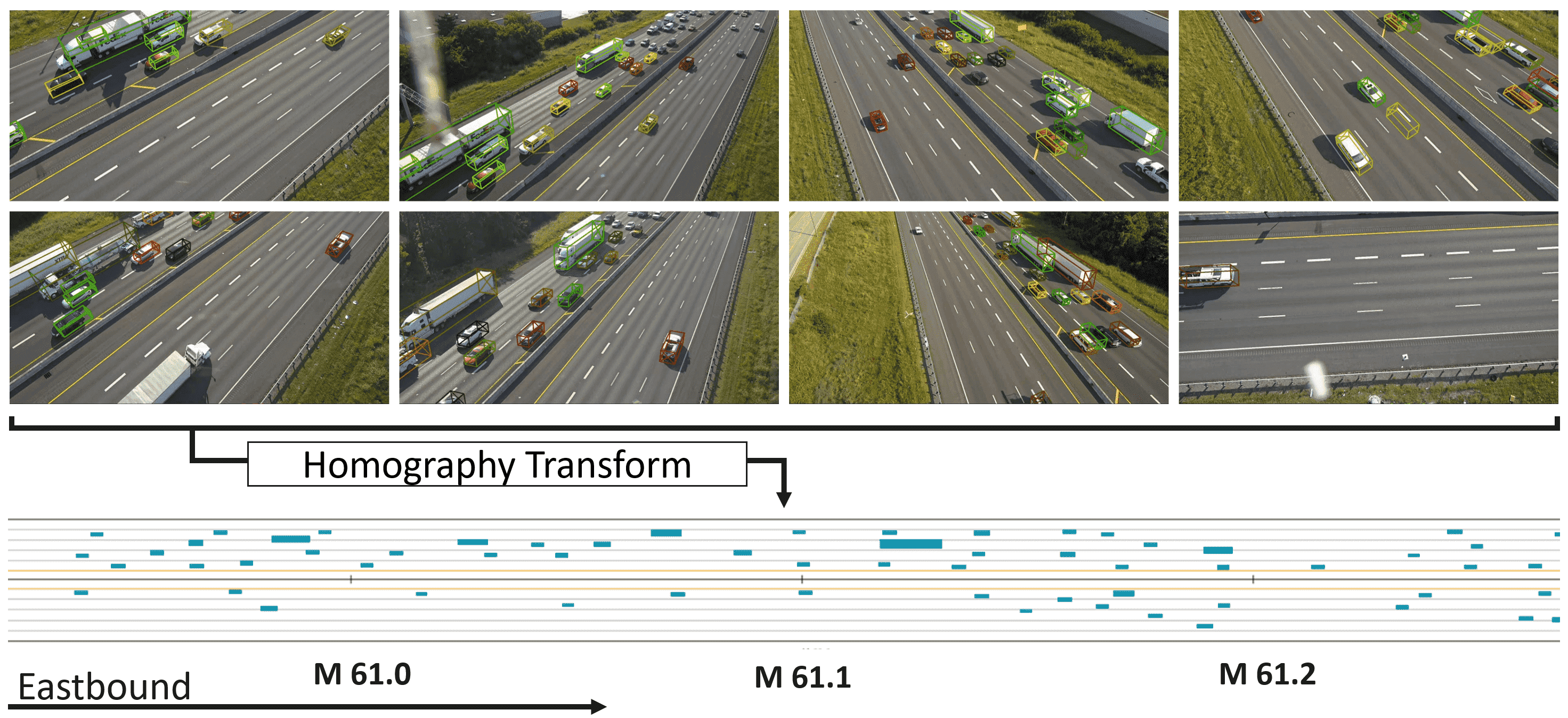}
    \caption{Vehicles are represented as 3D rectangular prism objects (various colors above) using object detection algorithms within each camera frame. The resulting detected objects are transformed into roadway coordinates shared among all cameras, and tracked in this unified coordinate system. Each blue rectangle in the projected 2D birds-eye view represents a vehicle position.}
    \label{fig:tracking}
\end{figure}

Vehicle detection and tracking is performed using \textit{Crop-based Tracking}, a joint detection and tracking method \cite{gloudemans2021vehicle}. This method processes only cropped portions of each overall image, drastically reducing detection inference time relative to processing each frame fully. Implicit in the use of this method is an accurate object motion model; object priors from this motion model are used to produce cropping boxes for each object, and only crops are processed by the object detector on most frames. For the base object detector, Retinanet with a ResNet-50 backbone is used \cite{Lin_2017_ICCV}. For the motion model, a Kalman filter with linear dynamics is used. Objects are assumed to travel with constant velocity along the primary direction of roadway travel, and are assumed to have zero velocity perpendicular to the primary roadway direction (note that this motion constraint is relaxed during data postprocessing and is only used during initial object tracking). The intersection-over-union metric is used to compute affinity between object positions and new detections \cite{bochinski2017high}. IOU is computed based on vehicle footprints in space rather than bounding box coordinates within an image, which allows detections from multiple cameras with distinct fields of view to be incorporated provided accurate homography information is available for each camera (for more information on camera homographies and data coordinate system, see Section \ref{sec:data-coordinates} and Appendix \ref{app:coord}. The multi-camera tracking problem is solved by detection fusion (as in \cite{strigel2013vehicle,luna2022online}) rather than trajectory fusion (as in \cite{wu2019multiview}) to reduce redundant tracking of the same object in multiple fields of view. Figure \ref{fig:tracking} shows the result of object detection and tracking within image coordinates, and the corresponding roadway coordinate object positions obtained using image homography.

The complete set of 276 camera fields of view is subdivided across multiple processing nodes. On each node, all cameras are processed together (that is, roughly one frame from each camera is processed at a time, subject to some frame skips to keep cameras tightly time-synchronized). Processing nodes are not synchronized, so a single object traveling through the full instrument extents will be tracked as a separate vehicle with a unique ID on each processing node.  This decouples the computation and allows the system to scale gracefully with a large number of cameras.

\subsubsection{Trajectory Post-processing}
Although raw trajectory data from dense deployment of cameras and CV algorithms can achieve complete spatial and temporal coverage of a roadway segment, such data contains inaccuracies from camera errors (dropped, doubled, and corrupted frames) network errors (data packet drops), object detection and tracking (fragmentations, ID swaps, false negatives and false positives \cite{bernardin2008evaluating}) often caused by object-object or infrastructure-object occlusions, timestamp quantization errors, homography assumption errors, and infeasible derivative quantities resulting from finite difference approximation over very short timescales. Treatments for specific sources of errors that rely on multiple iterations of rectification or require manual fine-tuning are not viable for longer term streaming datasets the I-24 MOTION is designed to produce. For small datasets, data cleaning and rectification with some manual involvement can address many common errors created in vehicular datasets~\cite{coifman2017critical}.

I-24 MOTION uses an automatic data post-processing pipeline~\cite{wang2022automatic} which will be continuously improved to automate as much of the data cleaning steps as possible. Currently, it consists of a) an online data association algorithm to solve a min-cost flow problem, which consequently matches fragments that belong to the same object, and b) a trajectory reconciliation algorithm, which is formulated as a quadratic program. This algorithm reconstructs realistic vehicle dynamics from disturbed detection data with trajectory derivative smoothing and outlier correction while minimally perturbing the original vehicle detections. The resulting trajectories automatically satisfy the internal consistency (differentiation of trajectories with speeds and accelerations). Future post-processing development will consider conflict resolution along with trajectory smoothing to produce feasible inter-vehicular distances for accurate microscopic traffic studies.

\section{Trajectory Data} 
\label{sec:data}
This section provides an overview of the data created by the I-24 MOTION system: its attributes, scale, conventions and coordinate system, known artifacts in the data, and a preliminary analysis of data accuracy. 

\subsection{Data Description}

One single continuous recording session on the I-24 MOTION instrument processed through the software pipeline (from Section~\ref{sec:software}) results in a vehicle trajectory dataset. Each dataset produced by the system consists of a collection of individual vehicle trajectories. An individual vehicle trajectory consists of vehicle attributes as well as motion information (see Table \ref{tab:data-attributes}). Trajectory positions record the 2D footprint of the back center of each car, and are re-sampled at a frequency of 25 Hz to allow exact timestamp-based indexing. Derivative quantities such as velocity, acceleration and steering angle can be directly computed with position information via, for example, finite difference. An example vehicle trajectory is included in Appendix~\ref{app:trajectory}. 

\begin{table}[!h]
    \centering
    \begin{tabular}{lccl}
        \toprule
         \textbf{Attribute} & \textbf{Type} & \textbf{Unit} & \textbf{Description} \\
         \midrule
         \textunderscore id             & 12-byte BSON  & $-$ & vehicle identifier unique across all I-24 MOTION data \\
         vehicle class      & int    & $-$ & \textit{0: sedan}, \textit{1: midsize}, \textit{2: pickup}, \textit{3: van}, \textit{4: semi}, \textit{5: truck}, \textit{6: motorcycle} \\
         first timestamp    & float     & s &  minimum unix timestamp for this trajectory \\
         last timestamp    & float      & s &  maximum unix timestamp for this trajectory \\
         timestamp          & [float]   & s & array of times at which vehicle positions are recorded \\
         x position         & [float]   & ft & array of longitudinal positions on roadway corresponding to each timestamp \\
         y position         & [float]   & ft & array of lateral positions on roadway corresponding to each timestamp \\
         starting x         & float     & ft   & longitudinal position on roadway at first timestamp \\
         ending x           & float     & ft   & longitudinal position on roadway at last timestamp \\
         length             & float     & ft   & vehicle length \\
         width             & float      & ft   & vehicle width \\
         height             & float     & ft   & vehicle height \\
         direction          & int       & $-$  & -1 if westbound, 1 if eastbound \\
         configuration ID   & int       & $-$ & identifier linking data to a unique metadata  indicating trajectory generation algorithm settings \\
         \bottomrule
         
    \end{tabular}
    \caption{Data attributes for a single vehicle trajectory. Square brackets indicate an array of values.}
    \label{tab:data-attributes}
\end{table}

Accompanying this work, 10 days of trajectory data are released from weekday morning traffic. Each dataset spans typically 4 hours, from 6:00 AM to 10:00 AM, covering morning rush hour conditions. (Data from Friday, November 25th instead covers 11 hours.) A variety of traffic conditions are present throughout the various days of data, including at least three crash-induced bottlenecks, one debris-induced bottleneck, high-traffic conditions with travelling waves, and free-flow traffic conditions. Table \ref{tab:dataset} summarizes the data released with this work. Additional metrics, statistics, time-space diagrams, and useful information can be found with the data release, as this information will change as the data is updated in future versions. Time-space diagrams for the westbound portion of the roadway on each day of trajectory data are included in Appendix \ref{app:ts}, and individual lane time-space diagrams for one day are shown in Appendix \ref{app:lane-ts}. Details on the data release are included in Section \ref{sec:data-avail}.

\input{dataset.tex}

\subsection{Data Coordinate System} \label{sec:data-coordinates} 
\begin{figure}[!ht]
    \centering
    \includegraphics[width = \textwidth]{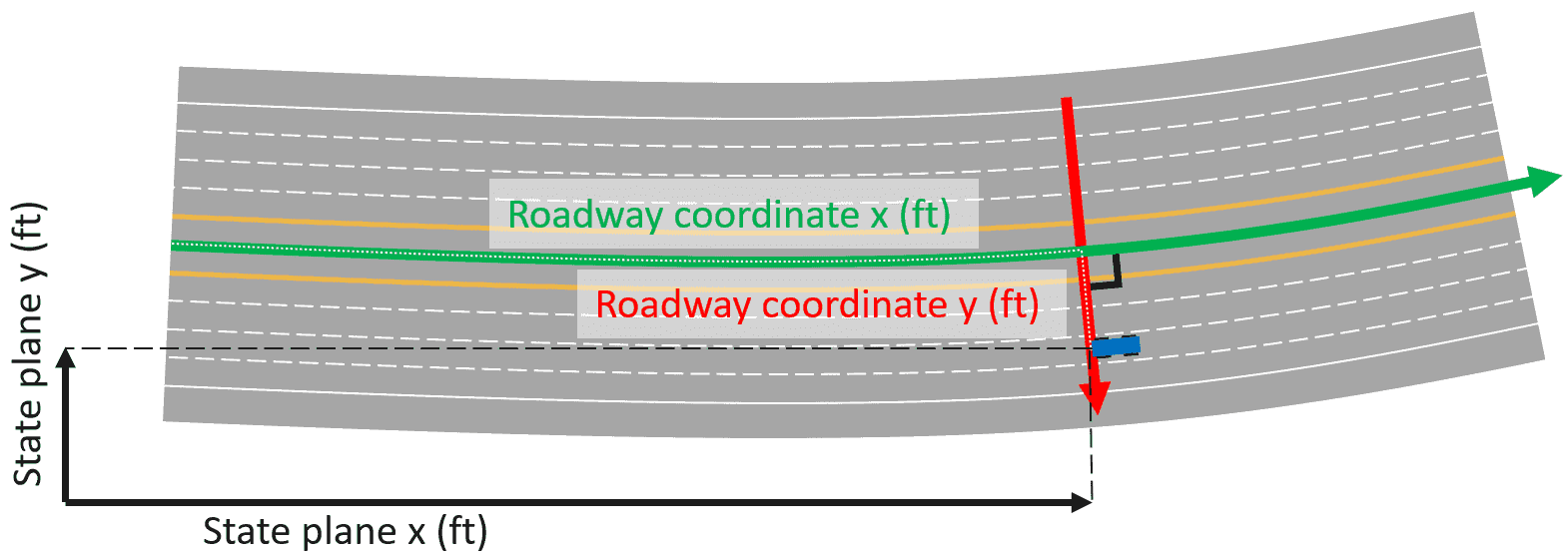}
    \caption{Spline-curvilinear $x$-axis (green) and locally perpendicular $y$-axis (red) for roadway coordinates. State plane coordinates are shown in black for comparison. Position of the vehicle can be expressed either in state plane coordinates (black dashes) or roadway coordinates (white dots).}
    \label{fig:coordinates}
\end{figure}

Data is provided natively in a curvilinear 2D roadway coordinate system, with the primary ($x$) axis aligned along the interstate roadway median and the secondary ($y$) axis defined locally perpendicular to the primary axis. This means that $x$ is roughly equivalent to station or mile marker along the roadway, while $y$ gives lateral or lane-position data. A second-order spline defines the $x$-axis in global (state plane) coordinates. (Control points for the center-line in state plane coordinates are included in metadata). This allows for the direct conversion of roadway coordinates into state plane coordinates, with a trivial conversion from state plane coordinates to GPS WGS84 coordinates. Both coordinate directions are stored natively in feet. The positive $x$-direction is defined in the eastbound direction (direction of increasing post-mile as defined by the Interstate 24 mile markers), and $x$-coordinates are offset such that the $x$-coordinate for post-mile 60 corresponds exactly to $5280 \times 60 = 316800$~ft. (Other postmiles are approximately but not exactly located in this way (e.g. post-mile 61 $\approx 5280 \times 61 = 322080$~ft.) Adopting the left-hand rule convention, the $y$-coordinate is positive on the eastbound side of the roadway (vehicle is moving in increasing $x$-direction). Figure \ref{fig:coordinates} illustrates the coordinate system, and Appendix \ref{app:coord} details the conversion of coordinates between the roadway coordinate system and state plane coordinates.

The primary advantages of a curvilinear coordinate system are twofold: \textit{i}.) The coordinate system aligns lateral (lane position) information along the $y$-axis, while accounting for the longitudinal curvature of the roadway and aligning the direction of travel with the $x$-axis. \textit{ii}.) A perpendicular slice of the roadway has a uniform $x$-coordinate.

While definition of the y-axis as locally perpendicular to the $x$-axis does allow for the same point to have multiple $(x,y)$ locations, for reasonable roadway curvatures these points occur suitably far from the roadway surface where the coordinate system is relevant. This coordinate system also slightly underestimates the distance travelled (and therefore the instantaneous speed) of vehicles on the exterior of a curve, relative to vehicles on the interior of a curve. the magnitude of this effect is no more than the ratio of roadway width to radius of curvature, which tends to be small (less than 5\%) on typical roadways. Exact distances travelled and speeds can instead be calculated by converting positions into state plane coordinates followed by finite difference calculation. 

\subsection{Positional Accuracy}
To assess the accuracy and suitability of I-24 MOTION trajectory data for micro-scale traffic analysis, output trajectory data is compared against an internal, manually labeled ground truth trajectory dataset, and onboard GPS information from instrumented vehicles traveling on the roadway.

\subsubsection{Manually Labeled Ground Truth}
Manual labeling of vehicles as 3D rectangular prism bounding boxes within videos from a subset of 18 cameras was performed for two scenarios: a free-flow traffic scenario and a highly congested (one side of roadway) scenario. In total, over 600,000 individual vehicle positions were labeled manually. The resulting vehicle trajectories were compared against the trajectory data output by running the I-24 MOTION trajectory generation algorithms on the same video data. For comparison, object positions were matched to \textit{ground truth} (GT) object positions as in \cite{berclaz2009multiple} at each timestep. A minimum \textit{intersection-over-union} (IOU) between the predicted and ground truth vehicle position was required to consider the predicted vehicle position a match for that ground truth object. Table \ref{tab:tracking_metrics} reports a number of multiple object tracking metrics for each scenario, as well as some metrics indicating the physical feasibility of the output trajectories. 97-98\% of ground truth objects have at least one predicted trajectory assigned to them (GT Match Rate) and for ground truth objects, on average 91-95\% of the overall trajectory is covered by matching predicted vehicle positions (Per GT Recall). Moreover, all vehicles have feasible accelerations, only 0-2\% of vehicle observations have infeasible heading angles, and only 0-2\% of vehicle trajectories overlap with another trajectory at some point. 

\input{table_metrics.tex}

\begin{figure*}[!htb]
    \centering
    \includegraphics[width = \textwidth]{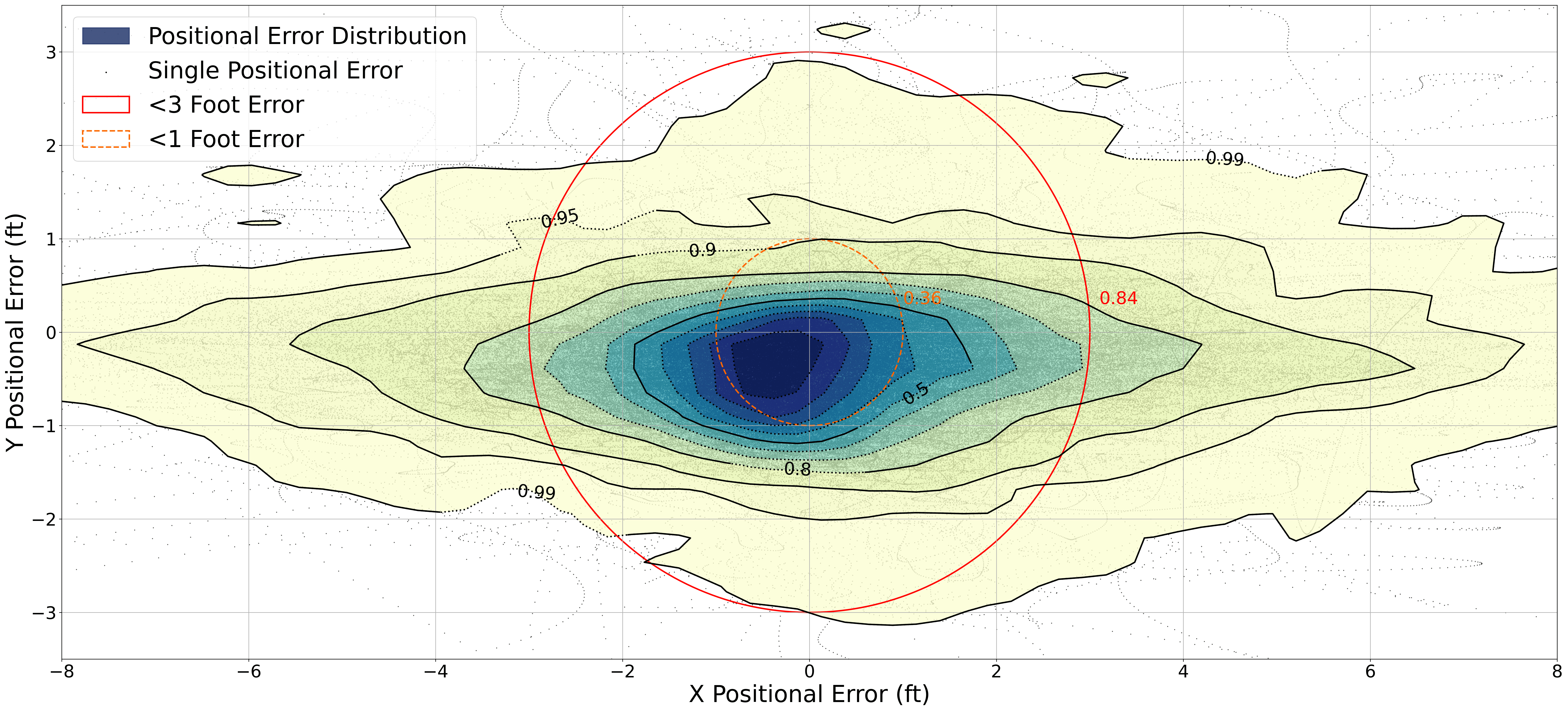}
    \caption{Positional error histogram for trajectory data relative to ground truth trajectories. Contours show the proportion of data is contained within, and are at intervals of 0.1 unless otherwise indicated. Single positional errors are shown as black dots. A red circle shows the proportion of data with less than 1-meter positional error (0.87) and an orange circle shows the proportion of data with less than 1-foot positional error (0.36).}
    \label{fig:bullseye}
\end{figure*}

For matched vehicle positions, Figure \ref{fig:bullseye} shows the relative error between the predicted and ground truth vehicle position. 84\% of predicted object positions fall within 3 feet of the ground truth position, and 36\% fall within 1 foot of the corresponding ground truth. Table \ref{tab:state_accuracy} reports the relative error between the predicted and ground truth vehicle dimensions. All dimensions have a mean absolute error of less than 1.2 feet.

\input{table_size.tex}

\subsubsection{GPS Data}
\begin{figure*}[!htb]
    \centering
    \includegraphics[width = 0.7\textwidth]{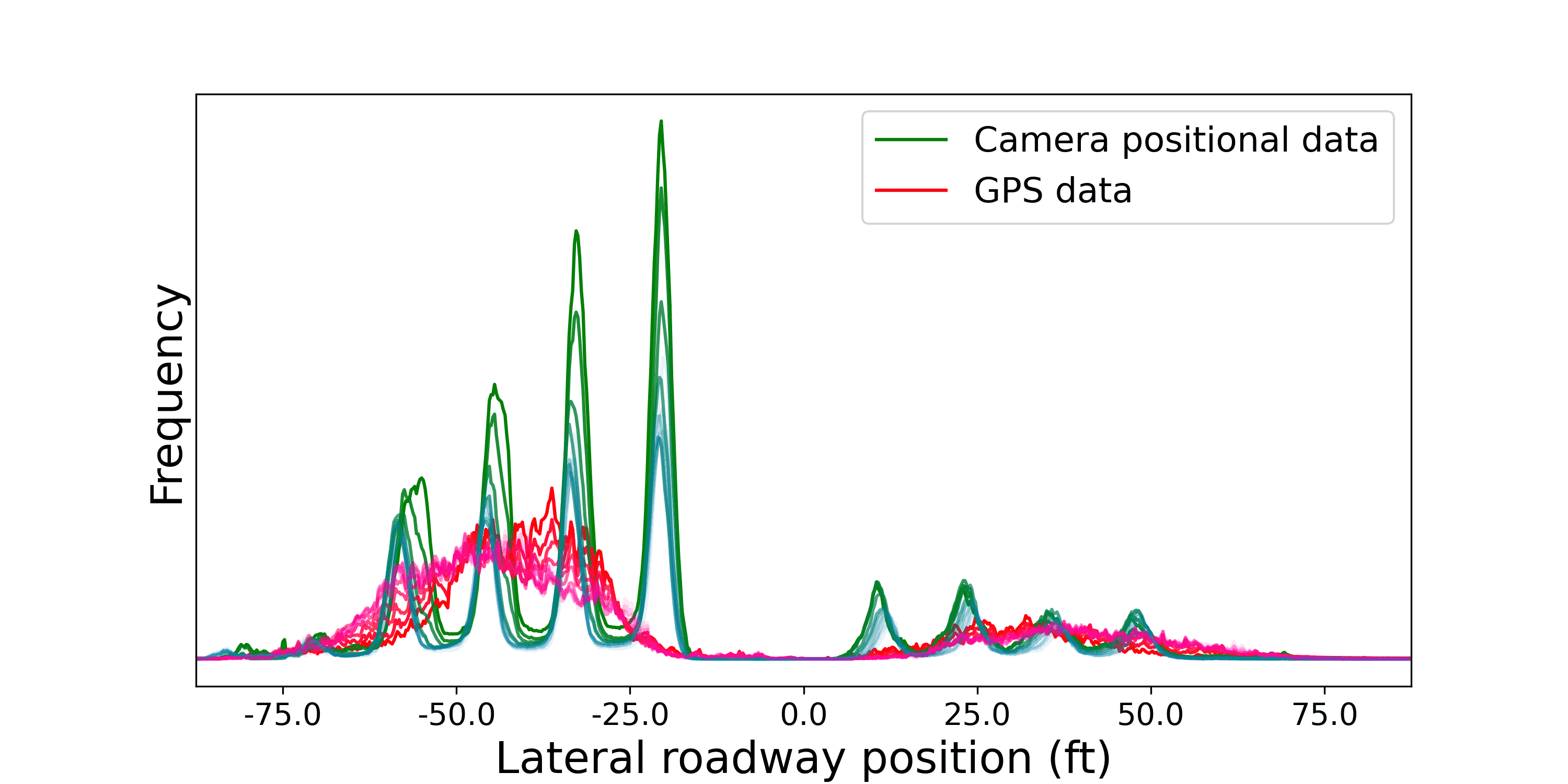}
    \caption{Lateral position histogram aggregated over several 1000-foot longitudinal slices, for I-24 MOTION camera trajectory data (blue-green) and onboard GPS data (pink-red). Strong peaks I-24 MOTION camera positional data correspond to lanes of travel. Data produced during AM rush-hour (higher traffic volume on westbound, negative lateral position, side of roadway).}
    \label{fig:gps}
\end{figure*}

Trajectory data was compared against onboard vehicle GPS sensor data, a commonly used sensor modality for obtaining single vehicle trajectories. GPS-equipped vehicles were driven in eastbound and westbound lanes of traffic on the I-24 MOTION instrument. Over 600 vehicle runs through the instrument extents were conducted. Regular (1 sec) positional data for each vehicle run was recorded. The reported \textit{circular error probable} (CEP) for the sensor was 2.5 meters. Figure \ref{fig:gps} shows a histogram of lateral positional data for each sensor modality (I-24 MOTION and GPS data), aggregated for several longitudinal slices along the instrument. The I-24 MOTION data shows strong lateral peaks corresponding to vehicle presence in a specific lane of travel, whereas the GPS lateral positional data does not show this characteristic. This is a strong indicator that I-24 MOTION yields strong lane-positional data, whereas this data is not necessarily available from an onboard GPS sensor without heavy filtering. 

\subsection{Data Artifacts} \label{sec:artifacts}
Relative to previous complete vehicle trajectory datasets, the data and instrument proposed in this work offer new challenges to perfect the data. Previous works were conducted in areas of sufficiently small spatio-temporal scale that physical occlusions could mostly be avoided (by overhead vantage point and careful roadway segment selection). Moreover, they were of sufficiently small temporal scale that errors remaining in the data after trajectory generation could be removed  with manual efforts \cite{coifman2017critical}). This approach is not scalable to the I-24 MOTION data, and some errors will always remain in the final data regardless of the algorithm employed. Enumerated here are a number of known errors in the initial data release that are artifacts of system hardware and software errors. We intend to partially or fully address each of these artifacts; moreover, open communication with I-24 MOTION data users will be maintained such that systematic errors in data creation can be addressed and data quality can be iteratively improved over time. 

\begin{figure} 
      \centering
      \includegraphics[width=\textwidth]{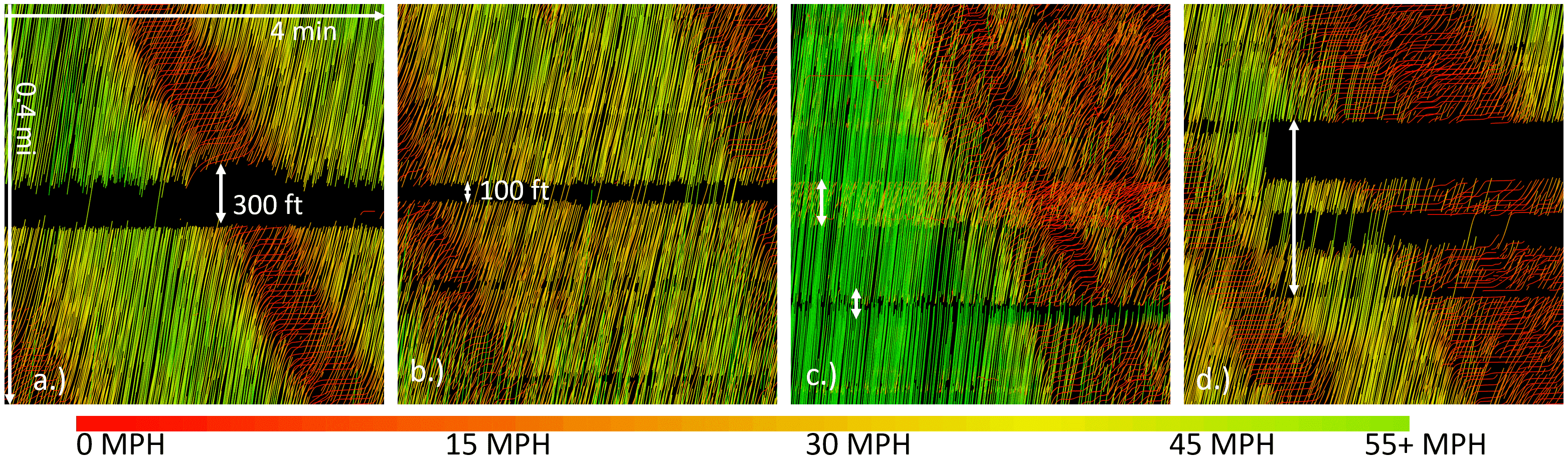}
      \caption{Example artifacts. For all figures, horizontal scale = 4 min. and vertical scale = 0.4 mi. a.) Missing pole causes a wide band of missing data. b.) Overpass causes a narrow band of missing data. In some cases post-processing can successfully stitch trajectories through this occlusion. c.) Homography error causes multiple trajectories corresponding to the same vehicle, or else results in a narrow band with no coverage. d.) Packet drops cause bands of missing trajectory data with a discrete start and end. Post-processing only partially fills in this data. }
      \label{fig:artifacts}
  \end{figure}

Figure \ref{fig:artifacts} shows time-space data with each type of data artifact present. Known data artifacts include:
\begin{itemize}
    \item \textbf{Missing Pole:} Data from a single pole is occasionally missing from one of two sources. Brief outages can occur due to network communication issues.  or to physical hardware damage (a camera pole was hit by a car in the week prior to most of the data in this work being generated). This manifests as a horizontal band on the time-space diagram (a contiguous spatial range of data missing across all recording time). Such issues are rare because poles are protected by guardrails, but these issues cannot be eliminated entirely.
    \item \textbf{Overpass Occlusion:} Overpass occlusion results in lost tracked vehicles, which also manifests as a contiguous spatial range of data missing across all recording time. This artifact will be addressed with an intelligent data processing step that matches objects disappearing under bridges with objects reappearing on the other side.
    \item \textbf{Static Homography Errors:} Initially, homographies for each camera were statically defined. However, pole deflection due to temperature and sunlight cause subtle shifts in camera positions. This manifests in very narrow (a few feet wide) horizontal bands on the time-space diagram that contain missing or doubled trajectory positional data. This issue will be corrected by periodically accounting for subtle camera motion by re-defining homographies.
    \item \textbf{Packet Drops / Frame Corruptions:} Network bandwidth limitations (especially near night-time hours when low light conditions create noisier and therefore larger video data) result in occasional packet drops or frame corruptions, which manifest as a band of missing positional data for a contiguous region of space and time. This issue will be mostly addressed by IP camera stream profile optimization and network connectivity improvements.
    \item \textbf{Fragmentations:} Ideally, each vehicle passing through the instrument is represented by a single recorded trajectory. In practice though, vehicles are often represented by several trajectory fragments, which are often the product of the above artifacts or other tracking or post-processing failures. Fragmentations manifest as discontinuous chunks of trajectory corresponding to a single vehicle. Fragmentations will be iteratively decreased over time as the above artifacts and other tracking issues are removed.
\end{itemize}

\subsection{Data Availability}  \label{sec:data-avail}
At the time of publication, data from I-24 MOTION will be made available on the project website located at \url{https://i24motion.org/data}. Data will be associated with a DOI for permanent referencing, and new versions of data will be assigned new DOIs according to standard DOI issuing guidelines. A README file contains information relevant to downloading, formatting, and using the data. Each processed day of data (a JSON set of JSON-like trajectories) is made available for download, as well as additional metadata including: scene homography for the data, trajectory extraction algorithm settings, and in-depth descriptions of data attributes. Data is initially released ``as is'', recognizing over time the data will be reprocessed and improved as the instrument matures.

Video data is in general not persistently recorded or made available with trajectory data. This is because the raw video data potentially contains \textit{personally identifiable information}. The instrument and data processing was designed to avoid collecting PII but it is difficult to guarantee no information was collected for all but very small subsets of data. We also note that the size of raw video files from the entire instrument is too large for easy distribution. For example the initial data release corresponds to approximately 47TB of video files.  Depending on research community needs and IRB considerations, it is possible this may be reconsidered in the future.

\section{Discussion}
  \label{sec:future}

This section provides some initial analysis of the datasets that are released with the publication. We generate the time-space diagrams of all of the published datasets, as well as illustrations of the type of analysis that can be conducted on the current data. 


\subsection{Traffic wave properties}
\begin{figure}
    \centering
    \begin{subfigure}{\linewidth}
    \centering
    \includegraphics[width=\linewidth]{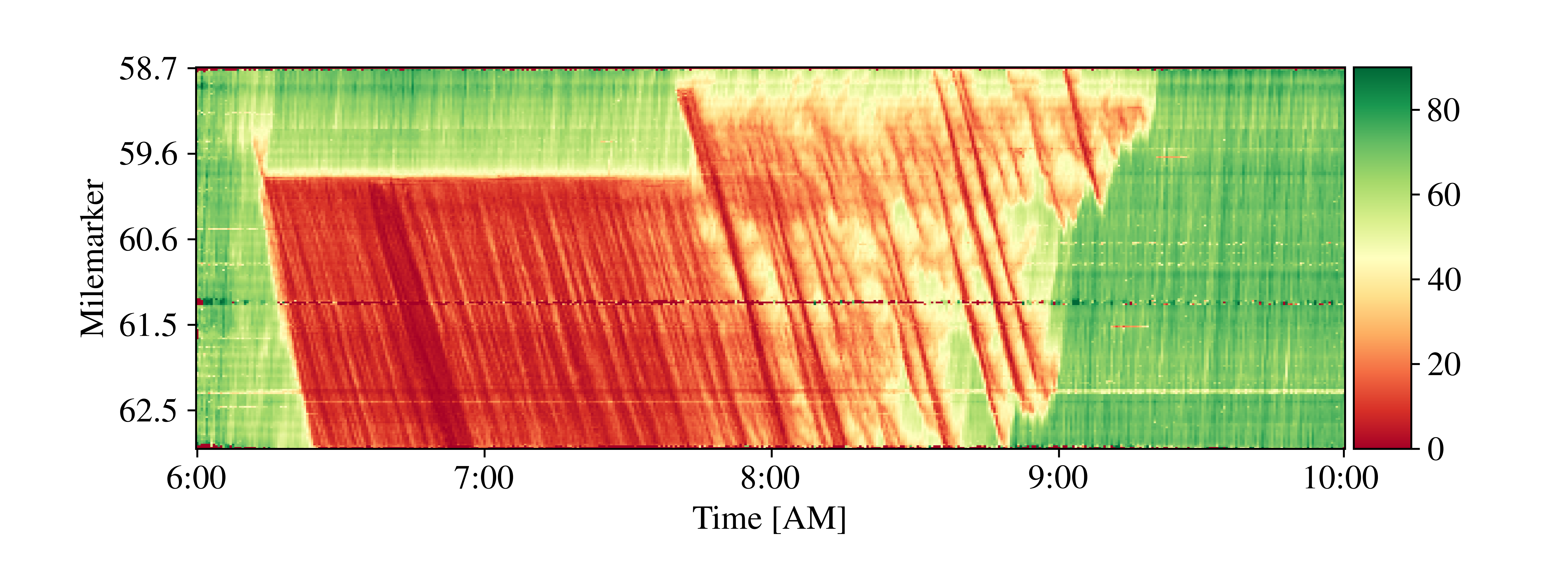}
    \caption{Monday Nov. 21 2022}
    \label{fig:heatmap_post1}
    \end{subfigure}
    \begin{subfigure}{\linewidth}
    \centering
    \includegraphics[width=\linewidth]{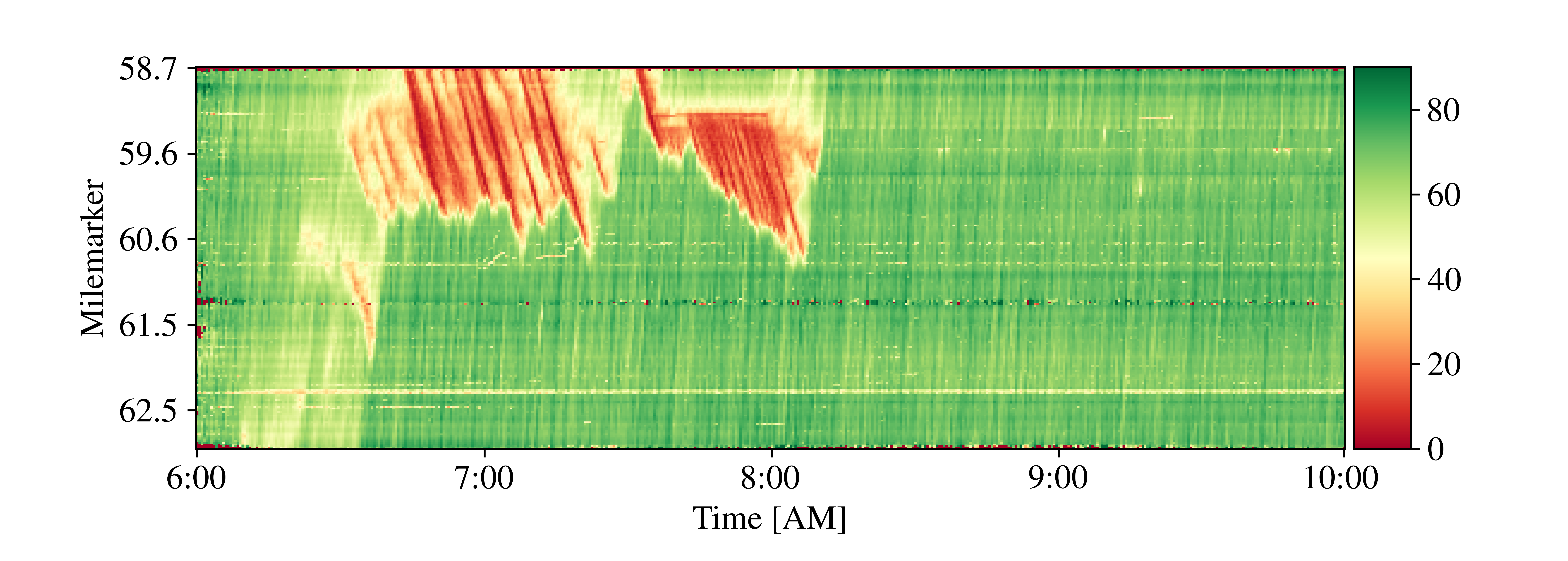}
        \caption{Wednesday Nov. 23 2022}
    \label{fig:heatmap_post3}
    \end{subfigure}
     \caption{Velocity field in (mph) obtained from the westbound (decreasing milemarkers) trajectory data on (a) Nov. 21 and (b) Nov. 23, 2022. Each plot depicts traffic velocity evolution during the morning rush hours on the 4-mile of I-24 MOTION main corridor. The velocity field is aggregated into small bins from trajectory data according to Edie's definitions~\cite{edie1963discussion} with grid size of $\Delta t=30$s and $\Delta x=100$ft, respectively. The window sizes are selected to preserve fine-scale traffic wave properties.
    }
     \label{fig:velocity_field}
  \end{figure}

  Traffic oscillations are characterized by regular acceleration/deceleration cycles in congested traffic, and is shown to have negative impact on the overall traffic efficiency and energy consumption~\cite{schonhof2007empirical,stern2018dissipation}. In this subsection we provide a few examples of macroscopic observations from a dataset captured by the I-24 MOTION system during the morning rush hours of two weekdays (Nov. 21 and Nov. 23, 2022) containing muiltiple events. The time space diagrams for these days are shown in Figure~\ref{fig:velocity_field}) including a variety of traffic patterns, such as free-flow, congested and stop-and-go traffic as well as bottlenecks caused by various incidents. 

  \input{table_wave.tex}
  We select three signature events from these  days (termed as Events A-C, see Table~\ref{tab:wave_properties}), which are incident-induced bottlenecks. Specifically, Event A is a severe rear-end crash on the HOV lane that was immediately followed by an onset of upstream queuing on lane 1 and lane 2. The congestion lasted for about 1.5 hrs before the crash was cleared. Event B is a slowdown on lane 3 caused by a large object falling out of a pickup truck. The roadway was cleared about 2.5 minutes later. Event C is a sideswipe crash due to a vehicle changing from lane 1 to lane 2 that caused a collision with another car travelling in lane 2. These events are summarized in Table~\ref{tab:wave_properties}.
 
  Characteristics of the waves upstream of the selected events are calculated and also summarized in Table~\ref{tab:wave_properties}, including the wave propagation speed, period (time it takes to experience a complete slowdown and speedup cycle at a fixed location), and amplitude (or fluctuation range). Here the wave property calculations are based on visual inspections combined with various well-known techniques such as wavelet transform~\cite{daubechies1992ten} and cross-correlation~\cite{Zielke2008}. We direct interested readers to common references such as~\cite{zheng2011applications,Zielke2008} for details.

  Figure~\ref{fig:velocity_field} shows that perturbations in different times and locations all propagate upstream. Although the periodicity and magnitude of the waves vary, depending on factors such as the severity of the bottleneck, road geometry, and heterogeneity of driver-vehicle units~\cite{Zielke2008}, they generally travel against the direction of traffic at a constant characteristic speed of approximately 13 mph (see also~\cite{treiber2010three, Helbing2009, kerner2005physics}). 
  We observe that oscillations with longer periods are often accompanied by larger amplitudes. For example, Event A has prominent waves with period 2.1 min and a speed range of 14.8 mph, Event B with period 5 min and a speed range of 34 mph, and Event C with period 1.8 min and a speed range of 10.8 mph, although the severity and the traffic conditions vary. The strong correlation between traffic wave period and amplitude is also discussed in~\cite{gartner2002traffic}. 

\begin{figure} 
      \centering
      \includegraphics[width=\textwidth]{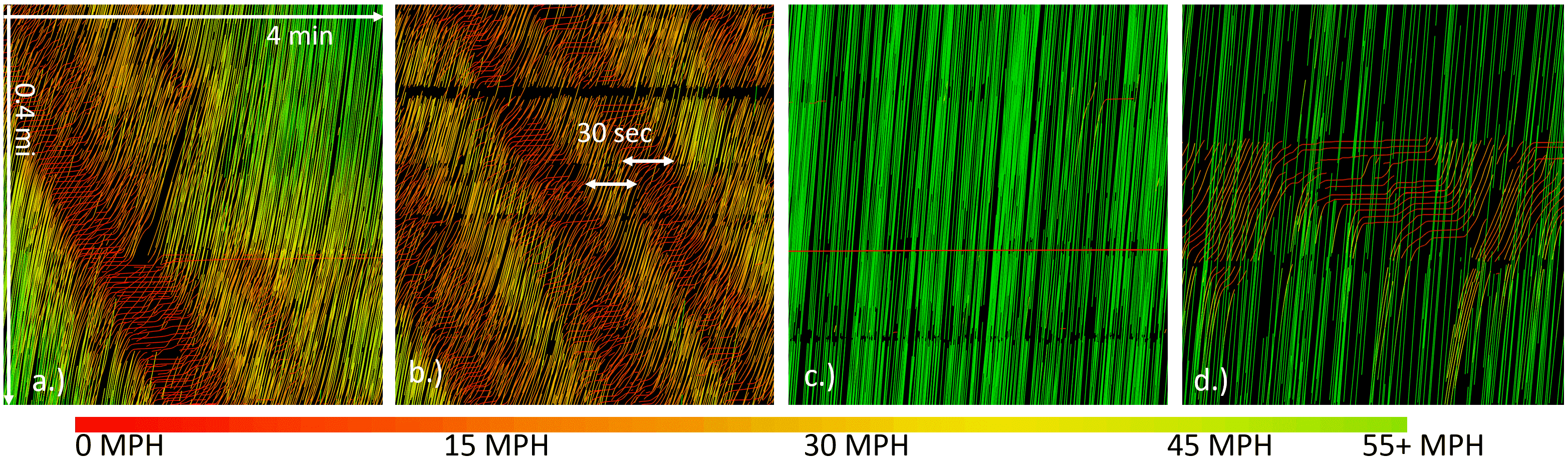}
      \caption{Examples of data phenomena difficult to observe in fixed-point or sparse GPS floating vehicle sensing schemes.  For all figures, horizontal scale = 4 min. and vertical scale = 0.4 mi. a.) Vehicle collision and resulting small-scale bottleneck. b.) Low-wavelength  ($\approx30$ sec) traffic waves in high-density flow. c.) A stopped vehicle on side of roadway. d.) Off-ramp queuing during otherwise free-flow conditions.}      \label{fig:phenomena}
  \end{figure}

  Even in the present form, data from I-24 MOTION already suitable to study traffic waves and other macroscopic quantities. This allows I-24 MOTION data to be used for speed analysis directly without needing to extrapolate long distances between fixed sensors (data cleaning is, however still required). Moreover, the camera-based sensors yield useful insight into the initial causes of bottlenecks not visible in any other sensing modality (e.g., debris on the roadway). Figure \ref{fig:phenomena} shows other example traffic phenomena not easily visible in traditional traffic sensing regimes.

\subsection{Fundamental diagrams (FDs)}
\begin{figure} [ht]
      \includegraphics[width=\linewidth]{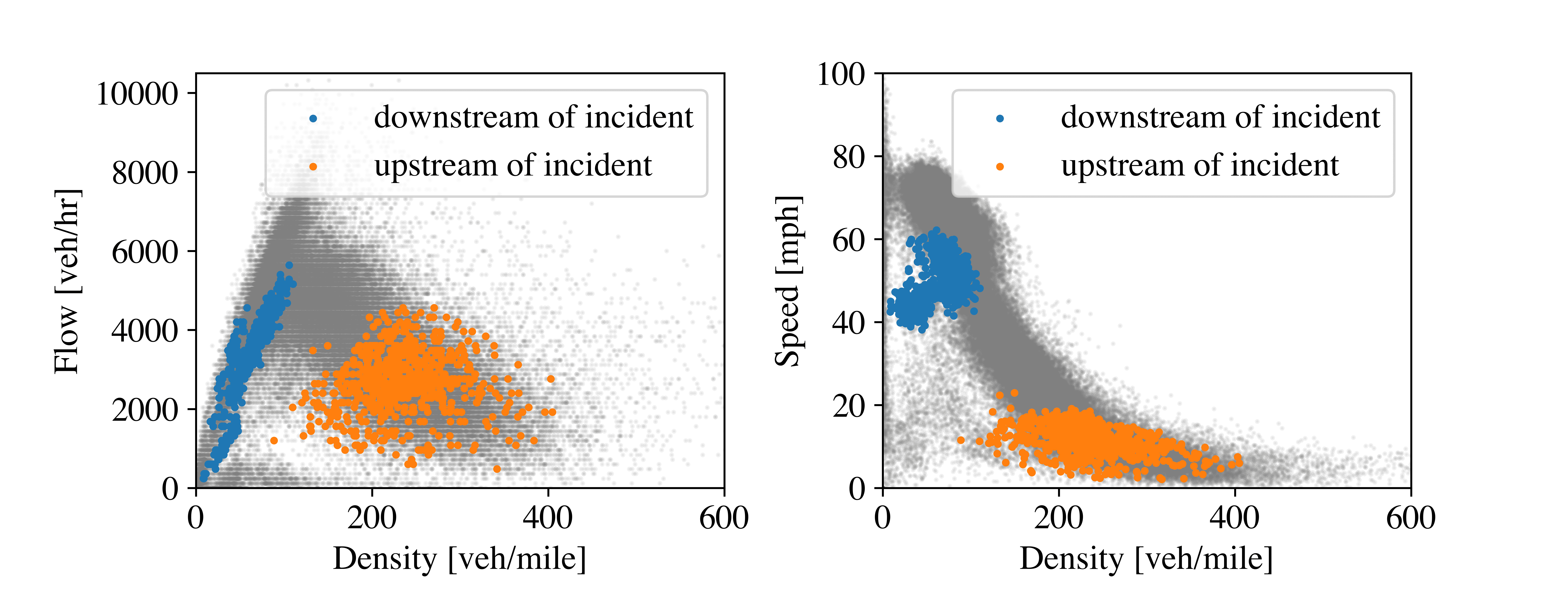}
      \caption{Traffic data immediately downstream and upstream of the crash on Monday Nov 21 (event A).}
      \label{fig:fd_post1}
  \end{figure}
  \begin{figure} [ht]
      \includegraphics[width=\linewidth]{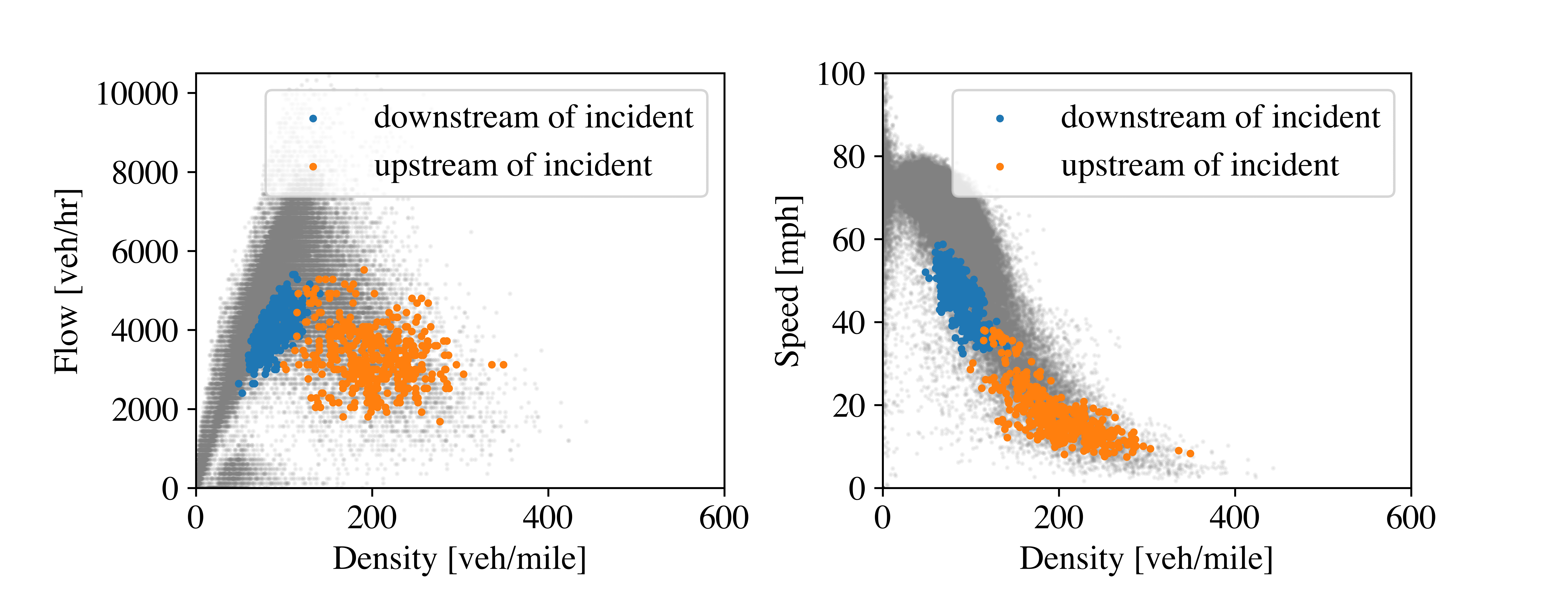}
      \caption{Traffic data immediately downstream and upstream of the crash on Wednesday Nov 23 (event C).}
      \label{fig:fd_post2}
  \end{figure}
  
  The empirical data from I-24 MOTION provides high resolution spatial-temporal evolution of traffic, which allows us to investigate more closely the changes of traffic properties on a finer scale. It also provides the possibility of computing fundamental diagrams at arbitrary locations around incidents. 
  
  For example, Figures~\ref{fig:fd_post1} and~\ref{fig:fd_post2} show fundamental diagrams and speed density plots computed immediately upstream and downstream of individual incidents, as well as the across the remainder of the dataset for the day. In each figure, the blue points correspond to speed/density/flow values downstream of the incident location, and the orange points correspond to the conditions immediately upstream of the crash at the same time. Data points shown in grey were collected at the same location on the day of the incident for reference. The points are computed from the trajectory data using Edie's definitions.  This illustrates a a capability that is possible to explore precisely because the complete roadway is monitored, allowing us to analyze the data around each event location.

The current illustrations provided here are not comprehensive but are rather designed to show that the data in its current form can already be used to support different research questions. As the datasets continue to improve, it will allow further investigations that bridge microscopic and macroscopic scales. 

\section{Conclusion}
\label{sec:conclusion}

This work introduces the I-24 MOTION instrument, which is designed to produce large scale trajectory datasets to support new directions in traffic science and traffic flow theory research. We also provide our initial datasets that will be improved and maintained as the instrument software continues to mature. 

Physical infrastructure construction on the instrument completed in November 2022, and the processing algorithms are far from final. In our ongoing work, we will be providing more datasets, tools, and methods, and software implementations that allow the instrument to support a wider range of applications and increase the overall data quality. Recognizing the evolving nature of the instrument, this work serves as a single reference to I-24 MOTION, with future works outlining the methodological improvements that advance data quality and provide insights into the traffic phenomena captured by the instrument. 

The instrument was also designed to support live experiments in traffic, including large deployments of automated vehicles which are designed to smooth traffic jams. The instrument will also support experiments conducted in collaboration with Tennessee Department of Transportation to support active traffic management, including experiments using variable speed limits, ramp meters, and lane closure systems. Such experiments will allow further investigation of the consequences of emerging technologies on traffic flow.

\clearpage
\pagebreak

\appendices

\section{I-24 MOTION Infrastructure Locations} \label{app:locations}
\input{table_gis.tex}

\clearpage
\pagebreak

\section{Example Vehicle Trajectory} \label{app:trajectory}

\input{table_example_trajectory.tex}
\newpage
\section{Coordinate System Conversion} \label{app:coord}
I-24 MOTION relies on 3 sets of coordinates:
\begin{itemize}
    \item \textbf{Image Coordinates:} are given in pixels. $(y^{(im)},x^{(im)})$ denotes the row and column of the specified pixel. By convention the top left pixel is (0,0).     
    \item \textbf{State Plane Coordinates:} specify a rectilinear and orthogonal coordinate system. The EPSG 2274 state plane coordinate system for Tennessee is specified in feet relative to a known survey point. $(x^{(st)},y^{(st)})$ indicates the coordinate (in feet) along the first (roughly horizontal) and second (roughly vertical) coordinate axis defined by the state plane coordinate system. (Note that a common conversion from state plane coordinates to latitude/longitude coordinates (e.g. WSG84 or NAD83) can be utilized if desired.) A third orthogonal coordinate axis (z-axis) is defined and corresponds to distance off the roadway, such that $z^{(st)} = 0$ for all points on the roadway plane. 
    \item \textbf{Roadway Coordinates:} are defined such that the primary (x) axis lies along the median (or more precisely, midway between the two interior yellow lines for the interstate) at all points within the instrument extents, and the secondary (y) axis is defined locally to perpendicular to the primary axis at all points along the roadway. Note that all coordinates with a distance from the primary axis less than the local radius of curvature have a unique $(x_r,y_r)$ coordinate. By left-hand rule convention, we  define the positive y-axis to be in the direction of the eastbound roadway lanes at all points along the roadway.
\end{itemize}

Throughout the rest of this appendix to disambiguate the various coordinate systems, the following notation is used: $x$,$y$, and $z$ refer to coordinate axes. A superscript $(im)$, $(st)$, or $(r)$  specifies all variables corresponding to a specific coordinate system (e.g. $x^{(st)}$). Vectors and matrices in that coordinate system are denoted in bold (e.g. $\textbf{O}^{(st)}$). Parameter matrices will be listed in ``mathcal'' script (e.g $\mathcal{H}$). A subscript indexes a specific point (e.g. $x^{(st)}_{bbl}$), and subscript $i$ indicates an arbitrary element index from a set of elements (e.g $a_i$). An $x$,$y$, or $z$ without a subscript indicates a generic variable along the specified axis within the specified coordinate system. A list of all variables along with their descriptions is given in Table \ref{tab:symbols}:

\begin{table}[ht]
    \centering
    \begin{tabular}{ll}
         \toprule
         \textbf{Symbol} & \textbf{Definition} \\
         \midrule
         
         $\mathcal{H}$          & 3$\times$3 matrix of homography parameters $h_{ij}$ \\ 
         $\mathcal{P}$          & 3$\times$4 matrix of homography parameters $p_{ij}$ \\ 
         $s$                    & homography scale parameter \\
         $x^{(im)}$, $y^{(im)}$ & image coordinates (y indicates pixel row and x indicates pixel column)\\
         $x^{(st)}$, $y^{(st)}$, $z^{(st)}$ & state plane coordinates \\
         $x^{(r)}$, $y^{(r)}$    & roadway coordinates \\
         $\mathbf{O}^{(st)}$     & state plane coordinates for object, equal to [$\mathbf{o}^{(st)}_{bbl}$,$\mathbf{o}^{(st)}_{bbr}$,$\mathbf{o}^{(st)}_{btl}$ ,$\mathbf{o}^{(st)}_{btr}$ ,$\mathbf{o}^{(st)}_{fbl}$ ,$\mathbf{o}^{(st)}_{fbr}$ ,$\mathbf{o}^{(st)}_{ftl}$ ,$\mathbf{o}^{(st)}_{ftr}$] \\ 
         
         $\mathbf{o}^{(st)}_{bbl}$ & back bottom left state plane coordinate of object, equal to [$x^{(st)}_{bbl},y^{(st)}_{bbl},z^{(st)}_{bbl}$] \\ 
         $\mathbf{o}^{(st)}_{c}$   & back bottom center state plane coordinate, primary reference coordinate for the object \\
         $\mathbf{o}^{(st)}_{spl}$ & state plane coordinates of point on center-line spline ($y^{(r)} = 0$) with the same $x^{(r)}$ coordinate as $\mathbf{o}^{(st)}_{c}$ \\

         $\mathbf{O}^{(r)}$ & roadway coordinates for object, $[x^{(r)_o},y^{(r)_o},l,w,h]$ \\
         $x^{(r)}$          & generic longitudinal roadway coordinate along curvilinear spline axis\\
         $y^{(r)}$          & generic lateral roadway coordinate along axis locally perpendicular to longitudinal roadway coordinate axis\\
         $x^{(r)}_o$          & object longitudinal roadway coordinate along curvilinear spline axis\\
         $y^{(r)}_o$          & object lateral roadway coordinate along axis locally perpendicular to longitudinal roadway coordinate axis\\
         $l$,$w$,$h$        & rectangular prism dimensions (length, width and height) \\
         $F(x^{(r)})$          & spline defining state plane coordinate roadway center-line spline parameterized by $x^{(r)}$ \\
         $\tilde{G}(x^{(st)})$ & spline approximating the  center-line spline in roadway coordinates $x^{(r)}$ parameterized by $x^{(st)}$ \\
         \bottomrule 
    \end{tabular}
    \caption{Summary of symbols used in Appendix \ref{app:coord}}
    \label{tab:symbols}
\end{table}

Transformations between image and state plane coordinates, and transformations between state plane and roadway coordinates are detailed in the next two sections.

\subsection{Image to State Plane Conversion}
A \textit{homography} relates two views of a planar surface. For each camera, we provide homography information such that the 8-corner coordinates of the stored 3D bounding-box annotation can be projected into any camera view for which the vehicle is visible, creating a monocular 3D bounding box within that camera field of view.  For each direction of travel in each camera view, for each scene, a homography relating the image pixel coordinates to the state plane coordinate system is defined. (Though the same cameras are used for different scenes, the positions of the cameras changes slightly over time due). A local flat plane assumption is used (the state plane coordinate system is assumed to be piece-wise flat) \cite{hartley2003multiple}. A series of correspondence points series of correspondence points $a_i = [x^{(im)},y^{(im)},x^{(st)},y^{(st)},z^{(st)}]$ are used to define this relation, where $(y^{(im)},x^{(im)})$ is the coordinate of selected correspondence point $a$ in pixel coordinates (row, column) and $(x^{(st)},y^{(st)},z^{(st)})$ is the selected correspondence point in state plane coordinates. 

All selected points are assumed to lie on the state plane, so $z^{(st)} = 0$ for all selected correspondence points. Visible lane marking lines and other easily recognizable landmarks on the roadway are used as correspondence points in each camera field of view. Each correspondence point is also labeled in \textit{global information system} (GIS) software, giving the precise GPS / state-plane coordinate system coordinates for each labeled corresponding point. The corresponding pixel coordinates are manually selected in each camera field of view, for each direction of travel on the roadway.

A \textit{perspective transform} (Equation \ref{eq:H}) is fit to these correspondence points. We first define a 2D perspective transform which defines a linear mapping (Equation \ref{eq:perspective}) of points from one plane to another that preserves straight lines. The correspondence points are then used to solve for the best perspective transform $\mathcal{H}$ as defined in equation \ref{eq:H}, where $s$ is a scale factor.

\begin{equation}
\label{eq:perspective}
    s \begin{bmatrix}
                x^{(st)}_i \\
                y^{(st)}_i \\
                1
        \end{bmatrix} 
        \sim
        \mathcal{H} 
            \begin{bmatrix}
                x^{(im)}_i \\
                y^{(im)}_i \\
                1
            \end{bmatrix}
\end{equation}

where $\mathcal{H}$ is a $3\times 3$ matrix of parameters:
\begin{equation}
\label{eq:H}
\mathcal{H} = 
            \begin{bmatrix}
            h_{11} & h_{12} & h_{13} \\
            h_{21} & h_{22} & h_{23} \\
            h_{31} & h_{32} & h_{33}
            \end{bmatrix} 
\end{equation}

 For each camera field of view and each direction of travel, the best perspective transform $\mathcal{H}^*$ is determined by minimizing the sum of squared re-projection errors according to equation \ref{eq:H_mse} as implemented in OpenCV's $find\mathunderscore homography$ function~\cite{bradski2000opencv}:
\begin{equation}
\label{eq:H_mse}
   \mathcal{H}^* = \argmin_\mathcal{H} \sum_i \left(x^{(st)}_i - \frac{h_{11}x^{(im)}_i + h_{12}y^{(im)}_i + h_{13}} {h_{31}x^{(im)}_i + h_{32}y^{(im)}_i + h_{33}}\right)^2 + \left(y^{(st)}_i - \frac{h_{21}x^{(im)}_i + h_{22}y^{(im)}_i + h_{23}} {h_{31}x^{(im)}_i + h_{32}y^{(im)}_i + h_{33}}\right)^2
\end{equation} 

The resulting matrix $\mathcal{H}^*$ allows any point lying on the plane within the camera field of view to be converted into state plane coordinates. The corresponding matrix $\mathcal{H}_{inv}$ can easily be obtained to convert roadway coordinates on the plane into image coordinates.  However, since each vehicle is represented by a 3D bounding box, the top corner coordinates of the box do not lie on the ground plane. A 3D perspective transform $\mathcal{P}$ is needed to linearly map coordinates from 3D state plane coordinate space to 2D image coordinate space, where $\mathcal{P}$ is a $3\times 4$ matrix of parameters:
\begin{equation}
\label{eq:P}
\mathcal{P} = 
            \begin{bmatrix}
            p_{11} & p_{12} & p_{13} & p_{14}\\
            p_{21} & p_{22} & p_{23} & p_{24}\\
            p_{31} & p_{32} & p_{33} & p_{34}
            \end{bmatrix} 
\end{equation}

and $\mathcal{P}$ projects a point in 3D space $(x',y',z')$ into the corresponding image point $(x,y)$ according to:

\begin{equation}
\label{eq:P_transform}
    \mathcal{P}  \begin{bmatrix}
                x^{(st)} \\
                y^{(st)} \\
                z^{(st)} \\
                1
        \end{bmatrix} 
        \sim
            s'
            \begin{bmatrix}
                x^{(im)} \\
                y^{(im)} \\
                1
            \end{bmatrix}
\end{equation}

where $s'$ is a new scaling parameter. By observing the case where $z^{(st)} = 0$, it is evident columns 1,2, and 4 of $\mathcal{P}$ are equivalent to the columns of $\mathcal{H}_{inv}$ and can be fit in the same way. Thus, we need only solve for column 3 of $\mathcal{P}$. Next, we note as in \cite{hartley2003multiple} that $(\frac{p_{11}}{p_{31}},\frac{p_{21}}{p_{31}})$ is the vanishing point (in image coordinates) of perspective lines drawn in the same direction as the state plane coordinate x-axis. The same is true for the 2nd column and the state plane coordinate y-axis, the 3rd column and the state plane coordinate z-axis, and the 4th column and the state plane coordinate origin.

Thus, to fully determine $\mathcal{P}$ it is sufficient to locate the vanishing point of the z-axis in state plane coordinates and to estimate the scaling parameter $p_{33}$. The vanishing point is located in image coordinates by finding the intersection point between lines drawn in the z-direction. Such lines are obtained by manually annotating vertical lines in each camera field of view. The scale parameter is estimated by minimizing the sum of squared reprojection errors defined in equation \ref{eq:P_mse} for a sufficiently large set of state plane coordinates and corresponding, manually annotated coordinates in image space.

\begin{align}
\label{eq:P_mse}
   \mathcal{P}^* = \argmin_{p_{33}} \sum_i &\left(x^{(im)}_i - \frac{p_{11}x^{(st)}_i + p_{12}y^{(st)}_i + p_{13}z^{(st)}_i + p_{14}} {p_{31}x^{(st)}_i + p_{32}y^{(st)}_i + p_{33}z^{(st)}_i + p_{34}}\right)^2 + \nonumber \\ &\left(y^{(im)}_i - \frac{p_{21}x^{(st)}_i + p_{22}y^{(st)}_i + p_{23}z^{(st)}_i + h_{24}} {p_{31}x^{(st)}_i + p_{32}y^{(st)}_i + p_{33}z^{(st)}_i + h_{34}}\right)^2
\end{align} 
 
 The resulting 3D perspective transform $\mathcal{P}^*$ allows for the lossless conversion of points in roadway coordinates to the corresponding points in image coordinates. Observing that a lossless conversion from image coordinates to state plane coordinates is available provided that the converted point lies on the $z^{(st)} = 0$ plane, it is possible to precisely convert a rectangular prism from image space to state plane coordinates by i.) converting the footprint of the prism near-losslessly into state plane coordinates (the only source of error comes from a set of 4 image coordinates that cannot be perfectly converted into a rectangle in state plane coordinates), ii.) shifting the footprint in state plane coordinates along the z-axis, iii.) re-projecting the resulting points back into the image, iv.) comparing the reprojected ``top points" to the original top of the rectangular prism in image coordinates, and v.) adjusting the height iterative to minimize the re-projection error until convergence. 

 \subsection{State Plane to Roadway Coordinate Conversion}

Next, we consider the conversion of points in state plane coordinates to roadway coordinates. In most cases, we care to convert a set of state plane coordinate points roughly in a rectangular prism (i.e. vehicle 3D bounding box) into roadway coordinates; thus, we define this conversion for a rectangular prism. A single point can be converted between state plane coordinates and roadway coordinates by treating it as a rectangular prism with zero length, width and height. 

Let $\mathbf{O}^{(st)}$ be a 3D bounding box representation in state plane coordinates, an 8$\times$3 matrix of x,y, and z coordinates for each corner of the box. (Note that these corners need not exactly correspond to an orthogonal rectangular prism, but the roadway coordinate equivalent will be exactly orthogonal so some truncation error will occur.) We reference, for example, the back bottom right (from the perspective of the rear of the vehicle) of object $\mathbf{O}^{(st)}$ as $\mathbf{o}_{bbr}^{(st)} = [x^{(st)}_{bbr},y^{(st)}_{bbr},z^{(st)}_{bbr}]$, such that  = $\mathbf{O}^{(st)} =[\mathbf{o}^{(st)}_{bbl},\mathbf{o}^{(st)}_{bbr},\mathbf{o}^{(st)}_{btl},\mathbf{o}^{(st)}_{btr},\mathbf{o}^{(st)}_{fbl},\mathbf{o}^{(st)}_{fbr},\mathbf{o}^{(st)}_{ftl},\mathbf{o}^{(st)}_{ftr}]$. (For the single-point case described above, all 8 corner coordinates are identical).

Next, Let $\mathbf{O}^{(r)} = [x^{(r)}_o,y^{(r)}_o,l,w,h]$ be the corresponding object representation of $\mathbf{O}^{(st)}$ in roadway coordinates. $x^{(r)}$ and $y^{(r)}$ are the roadway coordinate longitudinal and lateral coordinates (in feet), and $l$, $w$,and $h$ are the length, width, and height of the object respectively (in feet).

Let $\mathbf{o}^{(st)_c}$ denote the back bottom center coordinate of object $\mathbf{O}^{(st)}$. By convention, this point is referenced as the primary position of object $\mathbf{O}^{(st)}$. Let $\mathbf{o}^{(st)_{spl}}$ denote the point on the center-line spline (i.e. $y^{(r)} = 0$) with the same $x^{(r)}$ coordinate as $\mathbf{o}^{(st)}_c$.

Let $F$ be the  second-order spline parameterizing the roadway center-line in state plane coordinates. In other words, $F$ defines the longitudinal curvilinear axis $y^{(r)} =0$ along this spline. $F$ is fit by manually labeling a sufficiently large number of points along the interior yellow line for both directions of travel (in state plane coordinates). A spline is fit to each yellow line, and a third spline is fit to lie precisely halfway between these two splines. Spline control points are selected at suitably sparse intervals (200 foot minimum spacing) such that the spline is relatively smooth while still capturing the roadway curvature. 

Given $\mathbf{O}^{(st)}$, we first obtain $l$,$w$ and $h$ by computing the average distance between points on the front and back, left and right, or top and bottom of the vehicle respectively. Next, we obtain $\mathbf{o}^{(st)}_{c}$ by computing the average $x^{(st)}$ and $y^{(st)}$ state plane coordinates of the 4 back rectangular prism corners. 

Next, we solve for $x^{(r)}_o$ by solving the following optimization:

\begin{equation}
   x^{(r)}_o =  \argmin_{x^{(r)}} ( \text{dist}(F(x^{(r)}),\mathbf{o}^{(st)}_{c}))
\end{equation}

Where ``dist" indicates the Euclidean distance between the two points in state plane coordinate space. In other words, determine the point on the roadway spline closest to the back center of the rectangular prism $\mathbf{o}^{(st)}_{c}$. This minimizing point is the corresponding roadway longitudinal coordinate $x^{(r)}_o$, and the distance from the minimum distance point is roadway lateral coordinate $y^{(r)}_o$.

\begin{equation}
   y^{(r)}_o =  \min_{x^{(r)}} ( \text{dist}(F(x^{(r)}),\mathbf{o}^{(st)}_{c}))
\end{equation}

Noting that the I-24 MOTION roadway segment has monotonically increasing $x^{(st)}$ coordinate, we define a secondary spline $\tilde{G}(x^{(st)})$ to parameterize $x^{(r)}$ as a function of $x^{(st)}$, which yields a good initial guess for the closest roadway longitudinal coordinate for a given point in state plane coordinates. This optimization can then be solved to arbitrary precision, yielding the complete roadway coordinate for the object $\mathbf{O}^{(r)} = [x^{(r)}_o,y^{(r)}_o,l,w,h]$.

\subsection{Roadway to State Plane Coordinate Conversion}

Given roadway coordinates for an object $\mathbf{O}^{(r)}$, first find the corresponding point on the roadway center-line spline in state plane coordinates $\mathbf{o}^{(st)}_{c}$ according to:

\begin{equation}
     F(x^{(r)}_o)  = \mathbf{o}^{(st)}_{spl}
\end{equation}

To obtain the back center coordinate $\mathbf{o}^{(st)}_{c}$, we must offset $\mathbf{o}^{(st)}_{spl}$ by length $y^{(r)}$ in the direction perpendicular to the roadway centerline spline at $\mathbf{o}^{(st)}_{c}$. Let $\overrightarrow{\mathbf{u}}_{F}$ be the unit vector in the same direction as the derivative spline $F'$, and let $\overrightarrow{\mathbf{u}}_{1/F}$ be the unit vector in the perpendicular direction (along the state plane, i.e. $z^{(st)} = 0$. Note that care should be given to ensure that the positive direction of $\overrightarrow{\mathbf{u}}_{1/F}$ points towards the eastbound side of the roadway with positive $y^{(r)}$.) Then, $\mathbf{o}^{(st)}_{c}$ is given by: 

\begin{equation}
     \mathbf{o}^{(st)}_{c} = \mathbf{o}^{(st)}_{spl} + y^{(r)} \cdot \overrightarrow{\mathbf{u}}_{1/F}
\end{equation}

From here, the corner state plane coordinates for the right and left coordinates of the rectangular prism can be obtained by offsetting $\mathbf{o}^{(st)}_{c}$ by $ \pm \frac{1}{2}$ times $w$ in the direction of $\overrightarrow{\mathbf{u}}_{1/F}$, and the front coordinates of the rectangular prism can similarly be obtained by offsetting by $l$ in the direction of $\overrightarrow{\mathbf{u}}_{F}$ or in the opposite direction for  objects on the westbound or negative $y^{(r)}$ side of the roadway. Similarly, the top coordinates can be obtained by offsetting by a factor of $h$ in the $z^{(st)}$ direction. The direction of travel for an object can be obtained as the sign of the $y^{(r)}$ coordinate (negative for WB, positive for EB).For example, for an eastbound object the front top left coordinate can be obtained as:

\begin{equation}
    \mathbf{o}^{(st)}_{ftl}  = \mathbf{o}^{(st)}_{c} - \frac{1}{2} \cdot w \cdot \overrightarrow{\mathbf{u}}_{1/F} + l \cdot \overrightarrow{\mathbf{u}}_{F} + h \cdot \mathbf{[0,0,1]}
\end{equation}

\pagebreak

\section{Additional space time diagrams} \label{app:ts}
  \begin{figure} [!ht]
    \begin{subfigure}{\linewidth}
    \centering
    \includegraphics[width=\linewidth]{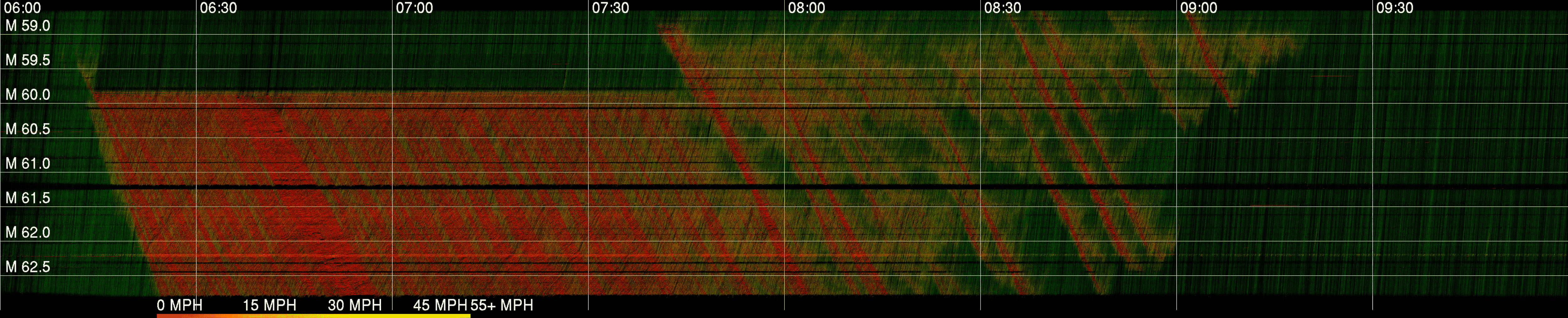}
    \caption{Monday Nov 21 2022}
    \label{fig:TS_2022-11-21}
    \end{subfigure}
    \begin{subfigure}{\linewidth}
    \centering
    \includegraphics[width=\linewidth]{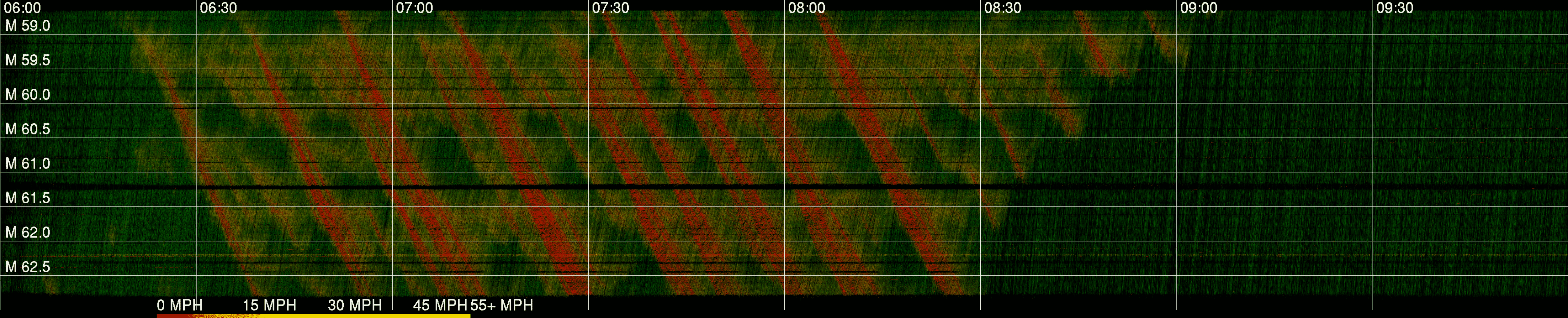}
        \caption{Tuesday Nov 22 2022}
    \label{fig:TS_2022-11-22}
    \end{subfigure}    
    \quad
    \begin{subfigure}{\linewidth}
    \centering
    \includegraphics[width=\linewidth]{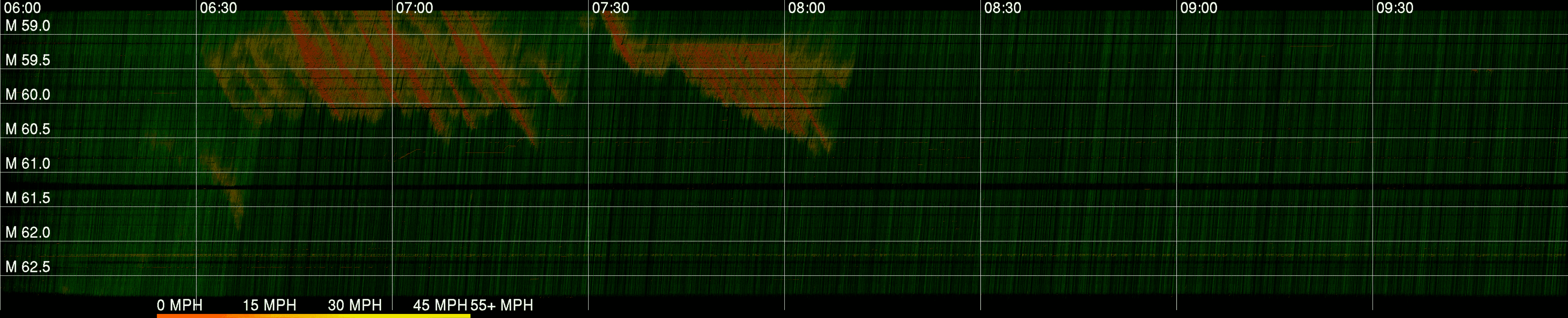}
        \caption{Wednesday Nov 23 2022}
    \label{fig:TS_2022-11-23}
    \end{subfigure}
    \begin{subfigure}{\linewidth}
    \centering
    \includegraphics[width=\linewidth]{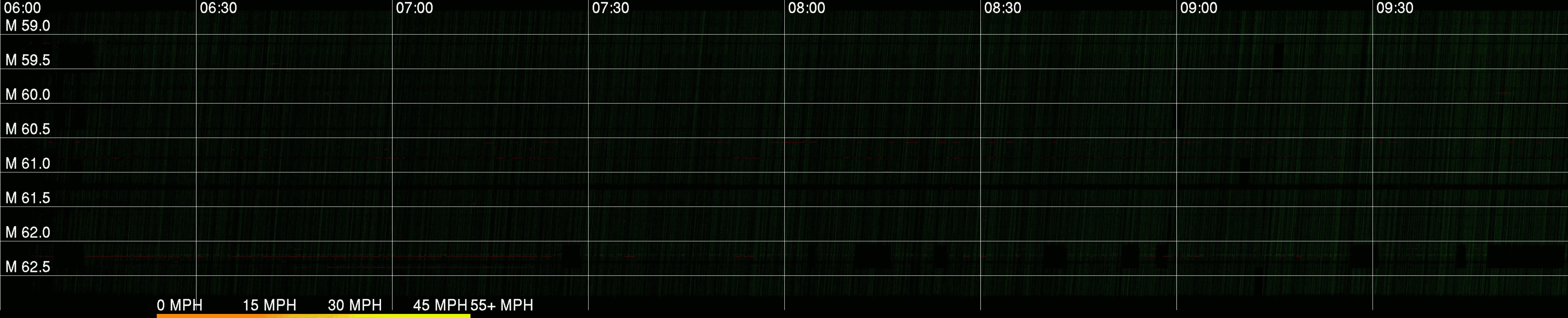}
        \caption{Thursday Nov 24 2022 (Thanksgiving)}
    \label{fig:TS_2022-11-24}
    \end{subfigure}
    \begin{subfigure}{\linewidth}
    \centering
    \includegraphics[width=\linewidth]{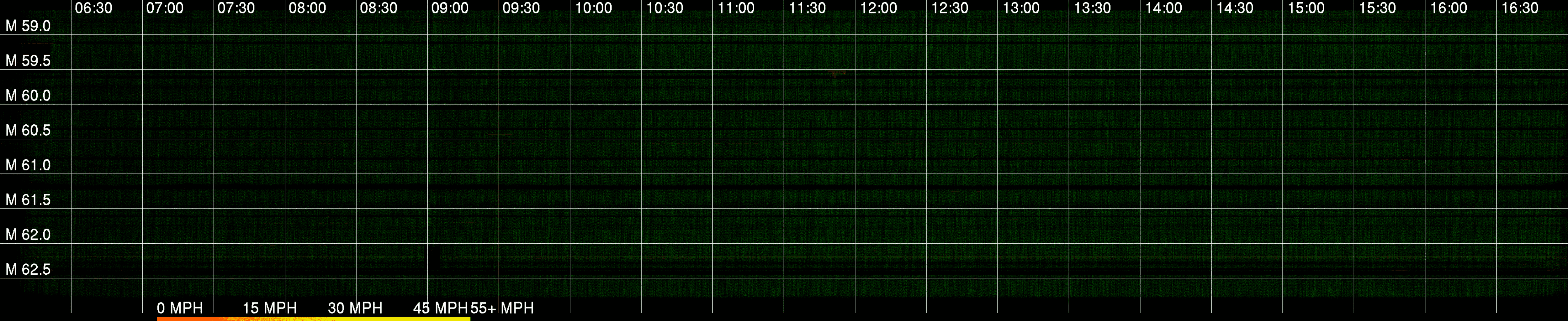}
        \caption{Friday Nov 25 2022 (Black Friday)}
    \label{fig:TS_2022-11-25}
    \end{subfigure}
    \quad
     
\end{figure}
\begin{figure}[!ht]
    \ContinuedFloat
    \begin{subfigure}{\linewidth}
    \centering
    \includegraphics[width=\linewidth]{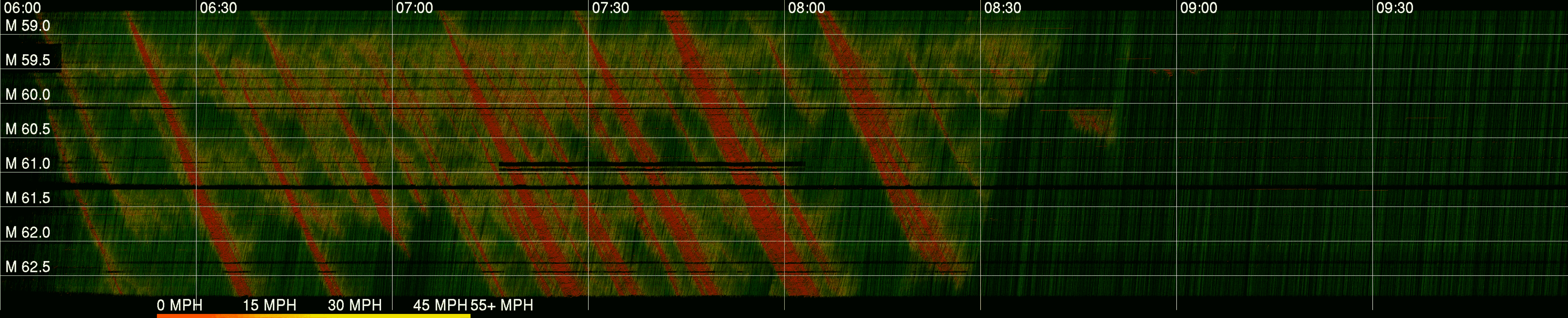}
        \caption{Monday Nov 28 2022}
    \label{fig:ST_2022-11-28}
    \end{subfigure}    
    \begin{subfigure}{\linewidth}
    \centering
    \includegraphics[width=\linewidth]{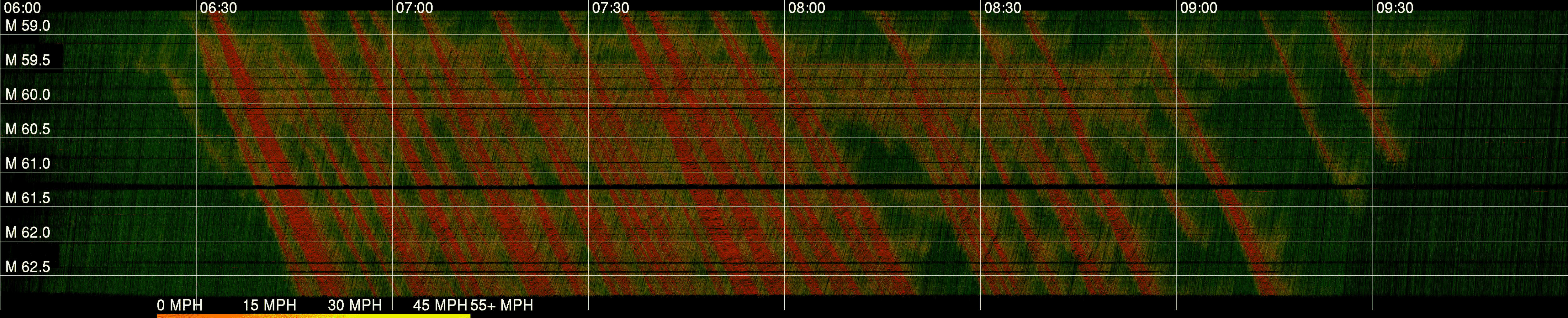}
        \caption{Tuesday Nov 29 2022}
    \label{fig:ST_2022-11-29}
    \end{subfigure}  
    \begin{subfigure}{\linewidth}
    \centering
    \includegraphics[width=\linewidth]{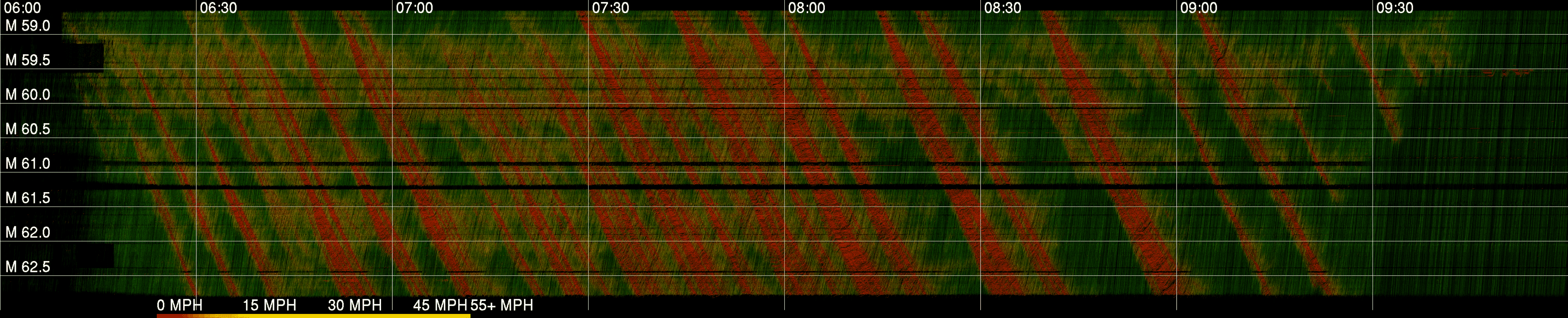}
        \caption{Wednesday Nov 30 2022}
    \label{fig:ST_2022-11-30}
    \end{subfigure}    
    \begin{subfigure}{\linewidth}
    \centering
    \includegraphics[width=\linewidth]{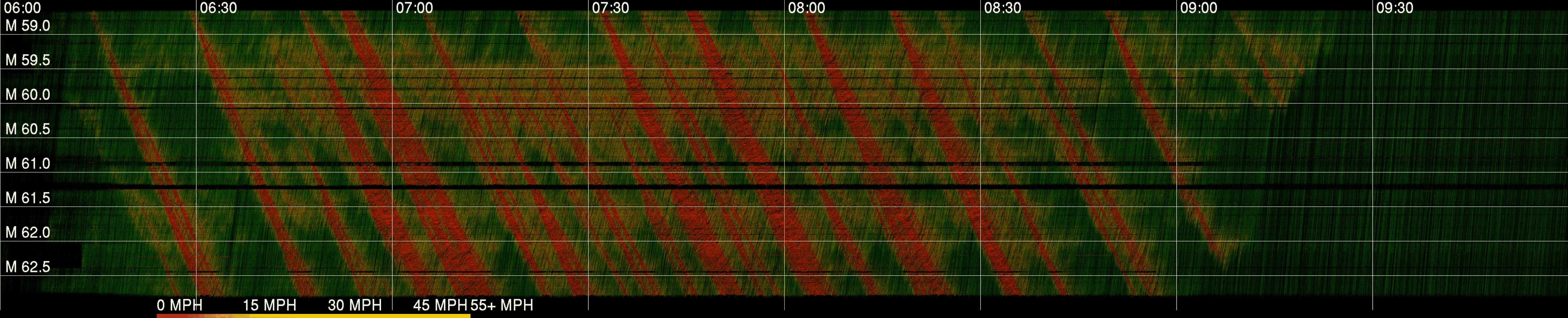}
        \caption{Thursday Dec 1 2022}
    \label{fig:ST_2022-12-01}
    \end{subfigure}  
    \begin{subfigure}{\linewidth}
    \centering
    \includegraphics[width=\linewidth]{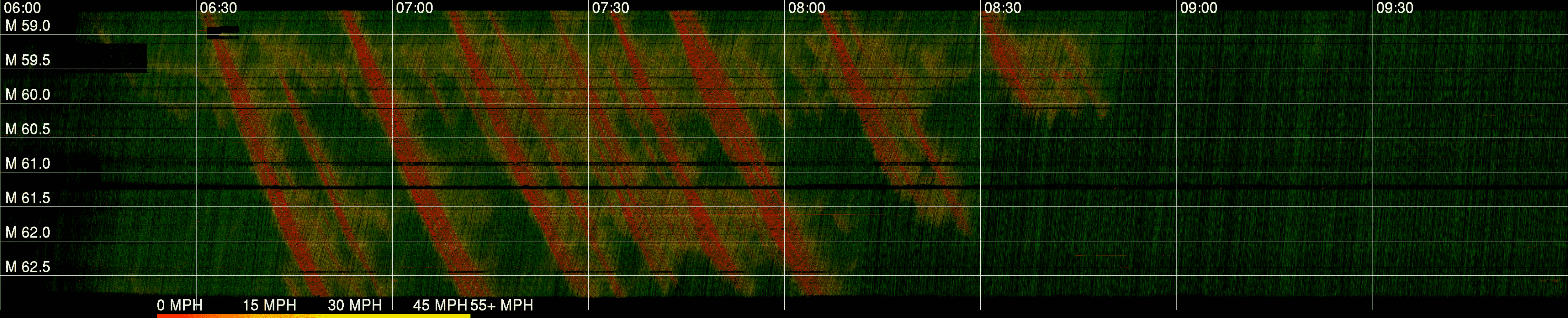}
        \caption{Friday Dec 2 2022}
    \label{fig:ST_2022-12-02}
    \end{subfigure}    
    \quad
    \caption{Additional time-space diagrams for I-24 westbound during morning rush hours on a) Nov 21, b) Nov 23, c) Nov 29, d) Dec 1 and e) Dec 2, 2022. }
    \label{fig:space_time_diagrams}
\end{figure}

\clearpage
\pagebreak

\section{Lane-Dis-aggregated Time-Space Diagrams} \label{app:lane-ts}
\begin{figure}[!ht]
    \centering
    \begin{subfigure}{\linewidth}
        \includegraphics[width=\linewidth]{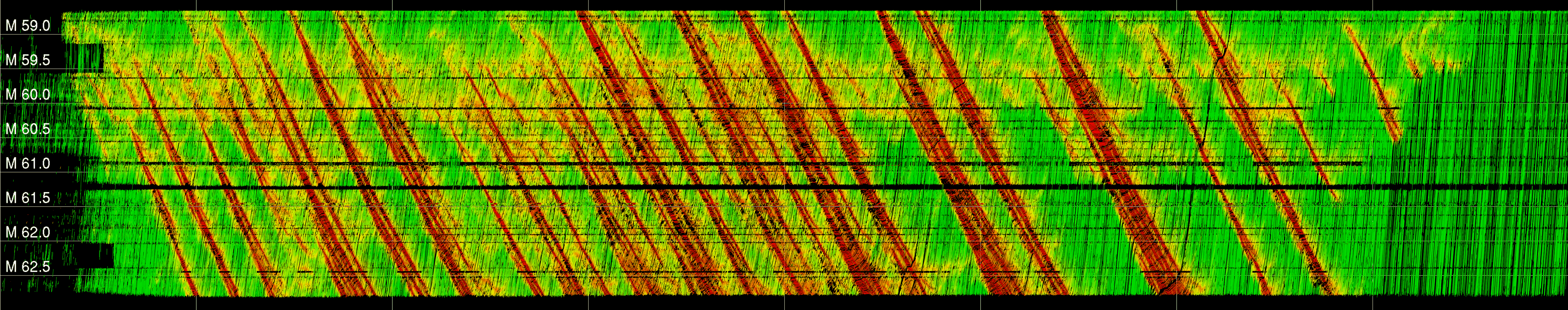}
        \caption{Lane 1 (HOV Lane)}        
    \end{subfigure}
    \begin{subfigure}{\linewidth}
        \includegraphics[width=\linewidth]{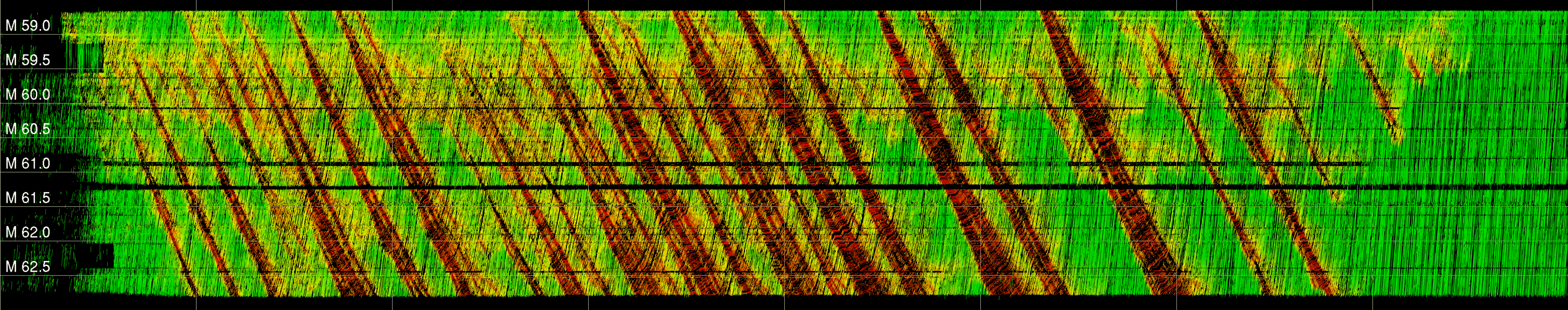}
        \caption{Lane 2}    
    \end{subfigure}
    \begin{subfigure}{\linewidth}
        \includegraphics[width=\linewidth]{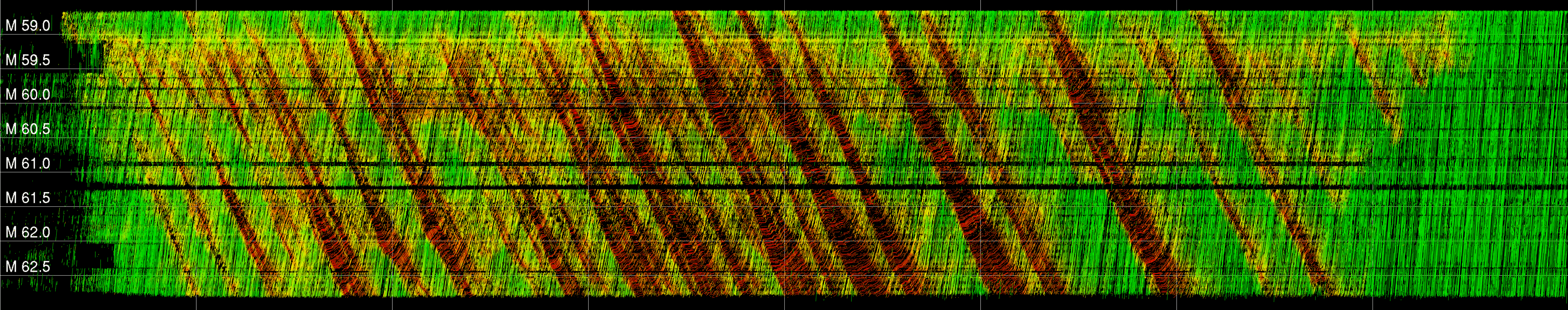}
        \caption{Lane 3}     
    \end{subfigure}
    \begin{subfigure}{\linewidth}
        \includegraphics[width=\linewidth]{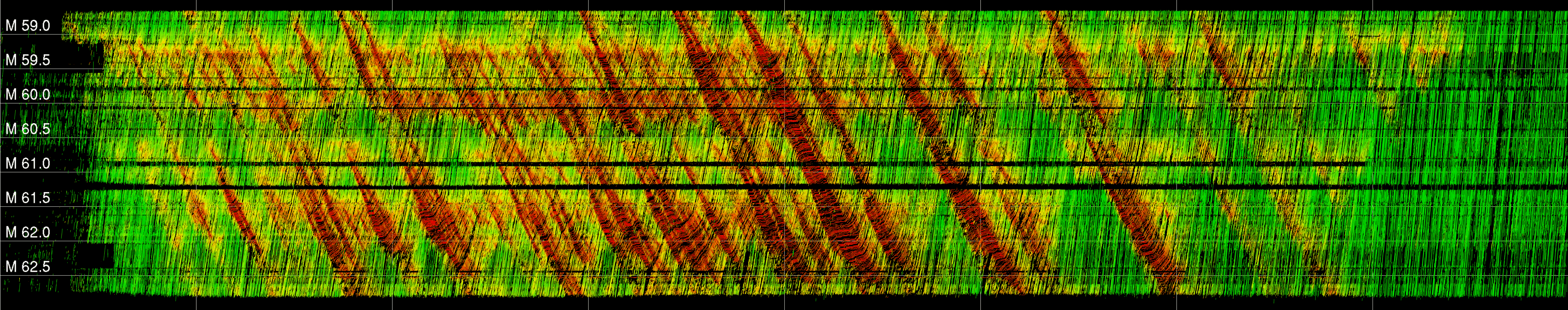}
        \caption{Lane 4}     
    \end{subfigure}    

    \begin{subfigure}{0.8\linewidth}
        \includegraphics[width=\linewidth]{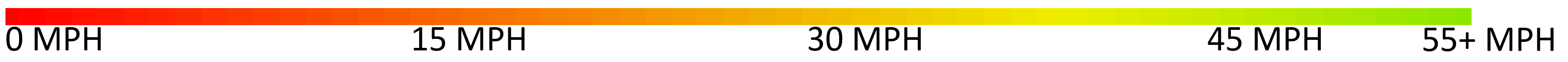}
    \end{subfigure} 
    
    \label{fig:lane_figures}
    \caption{Lane separated time-space diagrams (Wednesday, Nov 30 2022, 6:00-10:00 AM).}
\end{figure}

\clearpage

\section{Traffic Wave Calculations} \label{app:wave}
\subsection{Wave propagation speed}
The wave propagation speed is characterized by the slope of the slowdown that propagates upstream in the time-space diagram shown in~\ref{fig:velocity_field}. The slope is calculated based on the cross-correlation method as used in~\cite{coifman2005average,Zielke2008}, which compares the time series of the speed signals observed at two nearby locations on the same congested freeway. The idea is to shift one signal relative to another until the first non-trivial peaks are matched. The wave propagation speed is therefore the ratio between the time shifted and the distance of these two locations. We randomly select a few pairs of locations from one trajectory dataset and obtain a distribution of propagation speed. The distribution for the morning of Nov 22 2022, for example, has a mean of 12.8 mph and a standard deviation of 0.5 mph.

\subsection{Wave frequency analysis}
Wavelet transform is a time-frequency decomposition tool to effectively extract the non-stationary wave properties present in signals. The continuous wavelet transform is a convolution of the time-series signal $x(t)$ with a set of functions generated by the mother wavelet $\psi(t)$: 
\begin{equation}
    X_w(a, b)=\frac{1}{|a|^{1 / 2}} \int_{-\infty}^{\infty} x(t) {\psi}\left(\frac{t-b}{a}\right) d t,
\end{equation}
where $X_w(a, b)$ is a transformed signal at location $b$ and scale $a$ in the wavelet dimension. The scaling factor and the translation factor vary continuously, providing an overcomplete representation of the signals. We select a commonly used mother wavelet as a Morlet wavelet:
\begin{equation}
    \psi(t)=e ^{-\frac{t^2}{2}} \cos (5 t).
\end{equation}

\begin{figure}[!ht]
    \centering
    \includegraphics[width=0.8\linewidth]{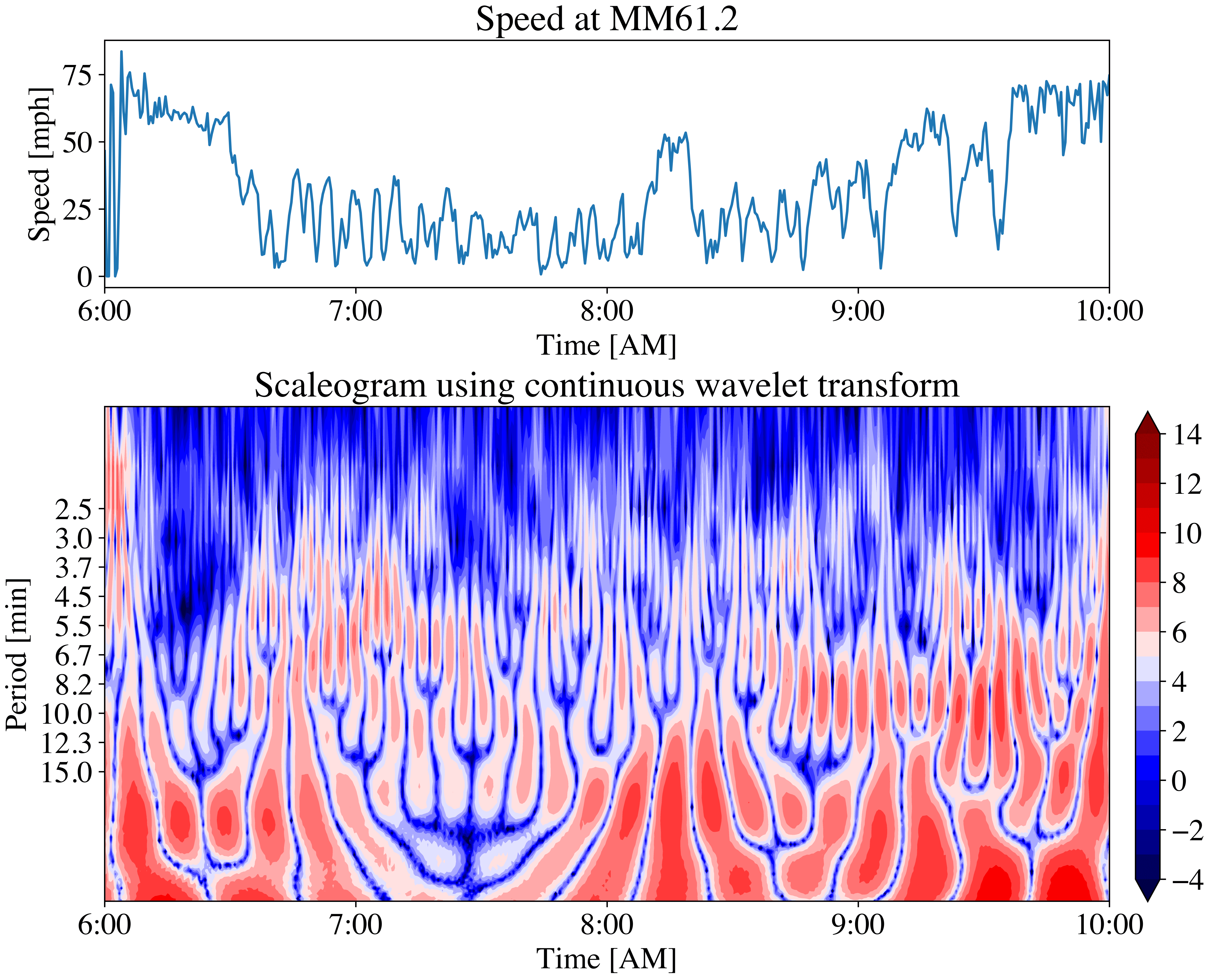}
    \caption{Top: the speed time-series sampled from MM61.2 on Tuesday, Nov 29 2022. Bottom: a scaleogram produced by continuous wavelet transform of the speed signal. The color represents log-scale of the power distribution across both frequency and time domain of the signal.}
    \label{fig:wt_firstcronjob}
\end{figure}


An example of wavelet transform result is shown in Figure~\ref{fig:wt_firstcronjob}. The top figure shows the time-series of speed sampled at a fixed location (in this case MM61.2) on Tuesday, Nov 29 2022. The bottom one is the corresponding wavelet transform scaleogram of the signal. It is obvious that the traffic waves do not appear to be stationary, i.e., the speed oscillation does not have a unique and consistent frequency across time. For example, during 6:50AM-7:30AM, the power of the signal peaks around 6.7min, corresponding to a salient wave period of 6.7min; during 8:30AM-9:30AM, the prominent wave period is near 9min. 

\section*{Acknowledgment}
The authors would like to thank Lee Smith, Brad Freeze, the Tennessee Department of Transportation, Meredith Cebelak, Matt D'Angelo and Gresham Smith for their efforts on conceptualizing, designing and implementing the system, Craig Philip and Janos Sztipanovits for their assistance conceptualizing I-24 MOTION. The authors would like to thank Eric Hall for his support on network, hardware, and software integration, and Zi Nean Teoh and Lisa Liu for their contributions to develop, build and deploy the I-24 MOTION system software architecture. The authors are grateful to Davis H. Elliot for constructing the instrument and WSP for serving as CEI for I-24 MOTION construction. This work is supported by the National Science Foundation (NSF) under Grant No. 2135579, the NSF Graduate Research Fellowship Grant No. DGE-1937963 and the USDOT Dwight D. Eisenhower Fellowship program under Grant No. 693JJ32245006 (Gloudemans) and No. 693JJ322NF5201 (Wang). This material is based upon work supported by the U.S. Department of Energy’s Office of Energy Efficiency and Renewable Energy (EERE) award number CID DE-EE0008872. The views expressed herein do not necessarily represent the views of the U.S. Department of Energy or the United States Government.

\ifCLASSOPTIONcaptionsoff
  \newpage
\fi



%
\bibliographystyle{IEEEtran}  
\bibliography{references}

\end{document}

%% file: dataset.tex
\begin{table}[!ht]
    \centering
    \begin{tabular}{lcccccl}
        \toprule
         \textbf{Date} & \textbf{Day} & \textbf{ID} & \textbf{Start time (AM)} & \textbf{Duration (hours)} & \textbf{Notes} \\
         \midrule
         Nov 21, 2022 & Monday & 637b023440527bf2daa5932f & 6:00 & 4    & crash, debris induced bottleneck \\ 
         Nov 22, 2022 & Tuesday & 637c399add50d54aa5af0cf4 & 6:00 & 4   & $-$ \\
         Nov 23, 2022 & Wednesday & 637d8ea678f0cb97981425dd & 6:00 & 4 &  crash \\
         Nov 24, 2022 & Thursday & 637f0d5f78f0cb97981425de & 6:00 & 4  & low traffic volume (holiday) \\
         Nov 25, 2022 & Friday & 6380728cdd50d54aa5af0cf5 & 6:00 & 11   &  low traffic volume (holiday) \\
         \midrule
         Nov 28, 2022 & Monday & 638450a3dd50d54aa5af0cf6 & 6:00 & 4    &  stopped vehicles induced slowdown \\
         Nov 29, 2022 & Tuesday & 63858a2cfb3ff533c12df166 & 6:00 & 4   &  $-$ \\
         Nov 30, 2022 & Wednesday & 6386d89efb3ff533c12df167 & 6:00 & 4 &  $-$ \\
         Dec 1, 2022 & Thursday & 63882be478f0cb97981425df & 6:00 & 4   &  merge induced slowdown \\
         Dec 2, 2022 & Friday & 63898d48d430891009401330 & 6:00 & 4     &  crash \\
         \bottomrule
    \end{tabular}
    \caption{Details of the released dataset. ``ID'' indicates the unique dataset identifier used to associate all data and metadata for this dataset. Additional summary and statistic information is included with the data release.}
    \label{tab:dataset}
\end{table}

%% file: table_metrics.tex
\begin{table*}[!ht]
\centering
\begin{tabular}{@{}lccl@{}}
    \toprule
    \textbf{Metric (1.0 best)}  & \textbf{Congested}& \textbf{Free-flow} & \textbf{Description} \\
    \midrule

    MOTA                 & 0.93 & 0.93 & Aggregate object tracking metric \\
    MOTP (IOU)           & 0.73 & 0.72 & Average precision (IOU) of matched object positions \\
    Precision            & 0.98 & 0.97 & Proportion of predicted object positions matched to a ground truth position \\
    Recall               & 0.95 & 0.96 & Proportion of ground truth object positions matched to a predicted object position \\
    GT Match Rate        & 0.97 & 0.98 & Proportion of ground truth trajectories matched to at least one predicted trajectory \\
    Pred Match Rate      & 0.99 & 0.76 & Proportion of predicted trajectories matched to at least one ground truth trajectory \\
    Per GT Recall        & 0.91 & 0.95 & Average proportion of a ground truth trajectory with correctly matched predicted object positions \\
    Per Pred Precision   & 0.98 & 0.74 & Average proportion of predicted trajectory correctly matched to a ground truth object \\
    Feas. Accel.          & 1.00 & 1.00 & Proportion of finite difference accelerations that are feasible ($ < 10ft/s^2$) \\
    Feas. Heading Angle  & 0.98 & 1.00 & Proportion of finite difference heading angles that are feasible ($ < 30 ^{\circ}$) \\
    Feas. Direction      & 0.99 & 1.00 & Proportion of finite difference velocities with correct magnitude (no backwards movement) \\
    Feas. Overlapping    & 0.98 & 1.00 & Proportion of predicted trajectories that never overlap with another trajectory \\
    \bottomrule
\end{tabular}
\caption{Multiple object tracking and trajectory feasibility metrics for two ground truth scenarios (congested and free flow).}
\label{tab:tracking_metrics}
\end{table*}

%% file: table_size.tex
\begin{table*}[!htb]
\centering
\begin{tabular}{@{}lccc@{}}
    \toprule
    \textbf{Quantity}  & \textbf{Mean Error (ft)} & \textbf{Standard Deviation(ft)} & \textbf{Mean Absolute Error (ft)}  \\
    \midrule

    Longitudinal (X) Position & 0.2 & 2.6 & 1.7 \\
    Lateral (Y) Position      & -0.3 & 0.6 & 0.6 \\
    Length                    & -0.6 & 2.5 & 1.2 \\
    Width                     & 0.1  & 0.5 & 0.3 \\
    Height                    & 0.5 & 0.8 & 0.7 \\
    \bottomrule
\end{tabular}
\caption{I-24 MOTION vehicle position and dimension errors relative to matched ground truth vehicles.}
\label{tab:state_accuracy}
\end{table*}

%% file: table_wave.tex
\begin{table*}[b]
\centering
\begin{tabular}{@{}cccccc|ccc@{}}
\toprule
\multicolumn{6}{c|}{\textbf{Event Information}}      & \multicolumn{3}{c}{\textbf{Upstream Wave Properties}}            \\ \midrule
\multicolumn{1}{c}{ Index} & \multicolumn{1}{c}{Date} & \multicolumn{1}{c}{Duration} & \multicolumn{1}{c}{\begin{tabular}[t]{@{}c@{}} Nearest \\ Milemarker\end{tabular}} & Description &\begin{tabular}[t]{@{}c@{}} Blocked \\ Lanes\end{tabular}& \multicolumn{1}{c}{\begin{tabular}[t]{@{}c@{}} Propagation \\ Speed (mph)\end{tabular}} & \multicolumn{1}{c}{\begin{tabular}[t]{@{}c@{}} Period \\ (min)\end{tabular}} & \multicolumn{1}{c}{\begin{tabular}[t]{@{}c@{}} Fluctuation \\ range (mph)\end{tabular}}\\ \midrule
                  A&      Nov 21&   6:14-7:43AM&      MM59.5&   Severe rear-end accident &1,2 and left shoulder &  12.6&  2.1&   0-14.8  \\
                  B&      Nov 21&   7:40-7:44AM&      MM58.8&   Debris in lane & 3  &  12.5&  5.0 &8.4-42.5\\
                  C&      Nov 23&   7:35-7:45AM&      MM59.5&   Sideswipe accident& 1 \& 2  &    13.1&  1.8 &  8.7-19.5 \\
                  \bottomrule
\end{tabular}
\caption{Approximate traffic wave properties in the upstream segment of selected events. The wave properties are obtained by a combination of  wavelet transform and visual inspection (see Appendix~\ref{app:wave}). Almost all waves appear to be ``quasi-periodic'' and non-stationary and therefore only the most prominent values are reported. }
\label{tab:wave_properties}
\end{table*}

%% file: table_gis.tex
\begin{table*}[ht]
\centering
\begin{tabular}{@{}ccc@{}}
    \toprule
    \textbf{Pole Number} & \textbf{Longitude}    & \textbf{Latitude}    \\
    \midrule
    1           & -86.6683396697044 & 36.0510246530758 \\
    2           & -86.6668725013732 & 36.0501702406015 \\
    3           & -86.6654402017593 & 36.0493331676102 \\
    4           & -86.6640186309814 & 36.0485112660384 \\
    5           & -86.6623315215110 & 36.0475679119687 \\
    6           & -86.6608375310897 & 36.0466917753056 \\
    7           & -86.6592979431152 & 36.0457939419758 \\
    8           & -86.6576912999153 & 36.0448483866264 \\
    9           & -86.6563770174980 & 36.0441218627600 \\
    10          & -86.6548401117324 & 36.0432196625972 \\
    11          & -86.6532951593399 & 36.0423391399701 \\
    12          & -86.6516268253326 & 36.0413740238091 \\
    13          & -86.6499182581901 & 36.0404305842180 \\
    14          & -86.6484859585762 & 36.0394762889946 \\
    15          & -86.6472119092941 & 36.0386521156311 \\
    16          & -86.6460505127906 & 36.0376630962090 \\
    17          & -86.6449534893035 & 36.0366393612687 \\
    18          & -86.6437357664108 & 36.0355310230575 \\
    19          & -86.6425502300262 & 36.0344812323649 \\
    20          & -86.6412305831909 & 36.0333142998429 \\
    21          & -86.6399243474006 & 36.0321169830776 \\
    22          & -86.6383659839630 & 36.0307396124762 \\
    23          & -86.6372233629226 & 36.0297092801383 \\
    24          & -86.6360297799110 & 36.0286377202112 \\
    25          & -86.6349676251411 & 36.0276897896641 \\
    26          & -86.6338652372360 & 36.0267006325845 \\
    27          & -86.6323256492614 & 36.0253340135559 \\
    28          & -86.6313627362251 & 36.0244836608630 \\
    29          & -86.6299653053283 & 36.0232428236304 \\
    30          & -86.6286617517471 & 36.0220887406933 \\
    31          & -86.6275808215141 & 36.0210930051775 \\
    32          & -86.6263175010681 & 36.0199736012082 \\
    33          & -86.6250273585319 & 36.0188173009549 \\
    34          & -86.6237318515777 & 36.0176848478644 \\
    35          & -86.6225999593734 & 36.0167042431535 \\
    36          & -86.6218060255050 & 36.0158603058791 \\
    37          & -86.6208162903785 & 36.0149816469298 \\
    38          & -86.6197273135185 & 36.0140248738225 \\
    39          & -86.6183325648307 & 36.0128055679068 \\
    40          & -86.6171014308929 & 36.0116643499086 \\
    Validation 1 & -86.6094464063644 & 36.0041353684958 \\
    Validation 2 & -86.6082850098609 & 36.0030981786895 \\
    Validation 3 & -86.6070619225502 & 36.0020240870002 \\
    \bottomrule
\end{tabular}
\caption{I-24 MOTION camera pole locations.}
\label{tab:poles}
\end{table*}

%% file: table_example_trajectory.tex

\begin{table}[ht]
    \centering
    \begin{tabular}{lccl}
        \toprule
         \textbf{Attribute} & \textbf{Type} & \textbf{Unit} & \textbf{Value} \\
         \midrule
         \textunderscore id                 & 12-byte BSON  & $-$ & 63732b74e1fa5a45ae0c2fdd \\
         vehicle class      & int    & $-$ & 0 \\
         first timestamp    & float     & s &  1668436223.30 \\
         last timestamp    & float     & s &  1668436257.60 \\
         timestamp          & [float]   & s & See Table \ref{tab:example} \\
         x position         & [float]   & ft & See Table \ref{tab:example} \\
         y position         & [float]   & ft & See Table \ref{tab:example} \\
         starting x         & float     & ft   & 325400.5531\\
         ending x           & float     & ft   & 329300.5458 \\
         length             & float     & ft   & 15.6381 \\
         width             & float      & ft   & 5.8521 \\
         height             & float     & ft   & 4.7021 \\
         direction          & int       & $-$  & 1 \\
         Configuration ID   & int       & $-$ & -1 \\
         \bottomrule
    \end{tabular}
    \caption{Detailed information of the example trajectory.}
    \label{tab:sample-static}
\end{table}

\begin{table*}[ht]
\centering
\begin{tabular}{ccc}
    \toprule
    \textbf{timestamp (s)}  & \textbf{x position (ft)} & \textbf{y position (ft)} \\
    \midrule
    1668436223.30  &  325400.5531 & -19.19265508 \\
    1668436223.34  &  325405.0238 & -19.12047988 \\
    1668436223.38  &  325409.4943 & -19.04921183 \\
    1668436223.42  &  325413.9646 & -18.97885093 \\
    1668436223.46  &  325418.4349 & -18.90939717 \\
     $\cdots$ & $\cdots$ & $\cdots$ \\
    1668436257.42  &  329281.8317 & -43.03453987 \\
    1668436257.46  &  329286.5097 & -43.09132499 \\
    1668436257.50  &  329291.1881 & -43.14893520 \\
    1668436257.54  &  329295.8668 & -43.20737050 \\
    1668436257.58  &  329300.5458 & -43.26663087 \\
    \bottomrule
\end{tabular}
\caption{The first 5 and the last 5 trajectory points for the example trajectory.}
\label{tab:example}
\end{table*}